\documentclass[conference,10pt]{IEEEtran}
\usepackage[english]{babel}
\usepackage{subfigure}
\usepackage{listings}
\usepackage{balance}
\usepackage{comment}
\usepackage{cite}
\usepackage{float}
\usepackage{mathtools,amssymb,amsfonts}
\usepackage{amsthm}
\usepackage{algorithmic}
\usepackage{graphicx}
\usepackage{textcomp}
\usepackage{multirow}
\usepackage{booktabs}
\usepackage[table]{xcolor}
\usepackage[draft]{todonotes}
\usepackage{cleveref}
\usepackage{xspace}
\usepackage{units}
\usepackage[hyphens]{url}
\usepackage{pifont}

\newcommand{\EC}{Edge Computing\xspace}

\newcommand{\CC}{Cloud Computing\xspace}
\newcommand{\WRT}{w.r.t.\xspace}
\newcommand{\percent}{\,\%\xspace}

\newcommand{\bpara}[1]{\noindent\textbf{#1}}

\definecolor{tablegray}{gray}{0.9}
\theoremstyle{definition}
\newtheorem{definition}{Definition}

\def\BibTeX{{\rm B\kern-.05em{\sc i\kern-.025em b}\kern-.08em
    T\kern-.1667em\lower.7ex\hbox{E}\kern-.125emX}}

\begin{document}

\author{\IEEEauthorblockN{Julien Gedeon, Martin Wagner, Karolis Skaisgiris, Florian Brandherm, Max M\"uhlh\"auser}
\IEEEauthorblockA{Telecooperation Lab, Technische Universit\"at Darmstadt, Germany\\
Email: \{gedeon, wagner, skaisgiris, brandherm, max\}@tk.tu-darmstadt.de}}
\title{Chameleons on Cloudlets: Elastic Edge Computing Through Microservice Variants}

\maketitle

\begin{abstract}

Common deployment models for \EC are based on (composable) microservices that are offloaded to cloudlets.
Runtime adaptations---in response to varying load, QoS fulfillment, mobility, etc.---are typically based on coarse-grained and costly management operations such as resource re-allocation or migration.
The services themselves, however, remain non-adaptive, worsening the already limited elasticity of \EC compared to \CC.
\EC applications often have stringent requirements on the execution time but are flexible regarding the quality of a computation.
The potential benefits of exploiting this trade-off remain untapped.
\par
This paper introduces the concept of \emph{adaptable microservices} that provide alternative variants of specific functionalities.
We define service variants that differ \WRT the \emph{internal} functioning of the service, manifested in different algorithms, parameters, and auxiliary data they use.
Such variants allow fine-grained trade-offs between the QoS (e.g., a maximum tolerable execution time) and the quality of the computation.
We integrate adaptable microservices into an \EC framework, show the practical impact of service variants, and present a strategy for switching variants at runtime.

\end{abstract}


\section{Introduction}
\label{sec:introduction}

\EC~\cite{gedeon2019-survey,Shi2016a,Satyanarayanan17-Emergence} is transforming our
computing landscape towards enabling low-latency computations at locations proximate to users.
Consequently, many platforms and frameworks exist that
propose \EC runtimes, e.g., built on lightweight container deployments
\cite{Pahl2015-Containers,Morabito2018a,Kaur2017-Energy,Liu2016-Paradrop}.
In addition, many \EC deployments follow the paradigm of microservices \cite{Filip2018-Microservices,Alam2018-Microservices}.
This abstraction has many advantages, such as higher agility and flexibility in the development of individual services.
At runtime, multiple microservices can be combined to form processing chains and, according to demands, individual services can be scaled in and out.
\par
While existing works are able to adapt the \emph{management} of the services, e.g., through the placement \cite{Salaht2020-ServicePlacement,Ouyang2018-FollowMeEdge}, or migration \cite{Wang2018-ServiceMigration,Ma2017-ServiceHandoff} of services, the services \emph{themselves} and their \emph{internal} functioning remain non-adaptable.
More specifically, edge services are implemented to deliver functionality in one particular way and cannot vary \emph{how} the functionality is provided, e.g., by providing different \emph{variants} of a service.
Those variants may, for instance, differ in the algorithms they use to perform a task.
As another example, some services need additional auxiliary data, which can also be varied (e.g., by using different pre-trained models for machine learning applications).
\par
Different variants of a microservice potentially have an impact on two metrics: (i)~the computational complexity, reflected in the execution time and resource demand of a request, and (ii)~the quality of result (QoR).
The latter can be defined and measured in different ways, e.g., by the accuracy of the result, i.e., its deviation from a (numeric) optimum, or by the (subjective) perception of a user.
Overall, these two metrics form a trade-off, in the sense that more accurate results typically require more computational effort, which leads to higher execution times and/or increased resource demands.
On the other hand, if we are willing to sacrifice computational quality, we can perform the same tasks with fewer resources.
\par
This observation is especially remarkable in the context of \EC if we recall some of its characteristics.
On the one hand, computing resources available in \EC are less powerful compared to their cloud counterparts, making efficient computing an important requirement to cope with scarce resources.
Similarly, achieving resource elasticity is more challenging in \EC, since the total available resources at a given location are much more limited.
On the other hand, many edge applications have stringent requirements on the overall latency.
At the same time, such mission-critical applications can be flexible regarding the quality of the computation result.
Examples can be found in the domain of image or video processing, and for recognition tasks.
To illustrate the practical impact of inaccurate computations, Chippa et al. \cite{Chippa2013-ACResilience} surveyed different kinds of applications and found that, on average, applications spent 83\percent of their execution time on computations that are error-tolerant.
Users of AR/VR headsets, for instance, might be willing to accept a lower rendering quality if in turn swift rendering helps in minimizing motion sickness.
Current \EC frameworks, however, do not consider this trade-off between computation effort and the quality of the computation result, and hence, miss out on this optimization opportunity.
\par
In this paper, we present the novel concept of \emph{adaptable microservices} for \EC.
We re-define microservices as blueprints for the delivery of a particular functionality that can be adapted \WRT (i)~the algorithms they use to perform a task, (ii)~parameters, and (iii)~auxiliary data required for the computation.
The possible \emph{variants} are implemented within the program code of a microservice and can be selected upon its instantiation.
Additionally, through a dedicated control channel, the current variant of a service instance can be changed at runtime.
The selection of the specific variant can be made according to certain requirements, e.g., a maximum tolerable execution time or a minimum quality of result.
Furthermore, by having service variants with varying resource requirements, service variants are a way to bring the much-valued resource elasticity of \CC to the domain of \EC.
\par
We argue that adapting service instances can be a way to avoid cold start latencies, as changes are applied to running services, instead of potentially re-starting a service or re-placing it, e.g., to more powerful hardware that can speed up its execution.
Service variations are applied to an individual microservice, but they also have to be considered in the context of a microservice chain.
For example, changing a variant of one microservice might have a disproportionate impact on the overall quality or execution time of the entire service chain.
\par
We propose including this concept of adaptable microservices in an \EC framework.
In such a system, clients submit an abstract definition of the desired microservice or service chain with their individual requirements regarding execution time and QoR to a controller, which in turn has to make the following decisions: (i)~which service variant to choose for instantiation in each step of the chain, and (ii)~the assignment of user requests to service instances (since multiple services in different variants might be available).
Furthermore, the controller might choose to change the variant of a particular microservice at runtime, e.g., by switching the algorithm with which the service performs its task.
Especially in cases where microservice instances are shared between multiple microservice chains, this becomes a non-trivial optimization problem, because users that share (parts of) a chain might have conflicting optimization goals.
\par
In summary, the contributions of this paper are threefold:
\begin{itemize}
  \item We propose the concept of \emph{adaptable microservices} in an \EC environment.
  We define the adaptability of microservices in three dimensions (algorithms, parameters, and auxiliary data).
  \item We present a concept for the integration of adaptable microservices into an \EC execution framework, detailing several components required for the orchestration of those service variants across edge cloudlets and users.
  \item We demonstrate the practical impact of orchestrating adaptable microservices \WRT (i)~profiling their execution time, (ii)~the effects of different variables on the service, and (iii)~how switching variants can adapt to changes in the request patterns.
\end{itemize}
The remainder of this paper is structured as follows. We review related work in \Cref{sec:backgroud_relatedwork}.
Our approach for dynamic microservice adaptation is presented in \Cref{sec:approach}.
\Cref{sec:implementation} details how we realize this concept in an \EC framework.
In \Cref{sec:evaluation}, we demonstrate the practical impact of microservice variants and their switching at runtime.
\Cref{sec:discussion_futurework} gives an outlook on future work before we conclude in \Cref{sec:conclusion}.

\section{Background \& Related Work}
\label{sec:backgroud_relatedwork}

Our contribution explores microservice adaptations in the context of \EC.
This section provides background information and reviews related work about \EC frameworks in general (\Cref{subsec:backgroud_relatedwork_edgeframeworks}), the concept of microservices and their adaptations (\Cref{subsec:backgroud_relatedwork_microservices_serverless}), and approximate computing (\Cref{sec:backgroud_relatedwork_approximatecomputing}).

\subsection{Edge Computing Frameworks}
\label{subsec:backgroud_relatedwork_edgeframeworks}

Since the emergence of \EC and the related concept of cloudlets \cite{Satyanarayanan2009-Cloudlets}, different implementations of this computing paradigm have been proposed \cite{Dolui2017-EdgeImplementations}.
For example, Carrega et al. \cite{Carrega2017-MEC} present a general middleware for Mobile \EC, that considers resources in the network of telecom operators.
The framework of Ferrer et al. \cite{Ferrer2019-AdHocEdge} focuses more on ad-hoc resources in the vicinity of users.
\par
Much of the research has been centered around management issues, such as the placement \cite{Salaht2020-ServicePlacement,Ouyang2018-FollowMeEdge,Wang2017-OnlinePlacement} of \EC services.
Furthermore, in dynamic \EC environments, migrations of users and services need to be handled accordingly \cite{Wang2018-ServiceMigration,Chen2019-ServiceMigration,Ma2017-ServiceHandoff}.
Some works like \emph{NOMAD}~\cite{Pamboris2015-NOMAD} integrate both caching and service migration strategies in an \EC platform.
\par
Similar to our implementation (see \Cref{subsec:implementation_framework}), some other works \cite{Liu2016-Paradrop, Mortazavi2017-Cloudpath, Bhardwaj2016-AirBox} also employ a repository for offloadable parts that are then transferred to surrogates.
Paradrop \cite{Liu2016-Paradrop} is an \EC platform that enables deploying third-party applications on WiFI access points.
\emph{CloudPath}~\cite{Mortazavi2017-Cloudpath} is restricted to stateless functions.
Both Paradrop and CloudPath do not allow the chaining of services.
Bhardwaj et al. \cite{Bhardwaj2016-AirBox} present \emph{AirBox}, a platform based on backend-driven onloading of functions to cloudlets.
Compared to our approach, none of these works allow adaptations to the \emph{internal} functioning of the provided services.
Wang et al. \cite{Wang2019-ScalableEdge} propose workload reduction as one method to increase scalability and elasticity for wearable cognitive assistance applications.
In comparison, our three dimensions of adaptation can be applied to a wider range of applications.

\subsection{Microservices, Service Adaptations, and Service Variants}
\label{subsec:backgroud_relatedwork_microservices_serverless}

Microservices are a contrasting paradigm to monolithic software.
Following the microservice paradigm, parts of an application are developed and deployed independently \cite{Fowler2012-Microservices}.
The benefits of microservices, e.g., regarding DevOps \cite{Balalaie2016-Microservices,Kang2016-DevOps} or scalability \cite{Jamshidi2018-Microservices,Dragoni2018-MicroserviceScalability} have been widely recognized.
However, developing applications as microservices also brings new challenges and concerns~\cite{Taibi2017-Microservices,Soldani2018-Microservices}, e.g., with regards to an increased operational complexity and testing efforts.
Dragoni et al. \cite{Dragoni2017-Microservices} provide a more in-depth introductory survey about the general concept of microservices.
Although the granularity of a microservice is not clearly defined \cite{Hassan2017-MicroserviceGranularity,Hassan2016-MicroserviceDesign}, microservices are typically characterized as small parts of an application with limited responsibilities, often restricted to performing a single task.
\par
Adaptation of services has been explored in the broader context of service-oriented architectures (SOA) and Web Services (WS) \cite{Papazoglou2003-SOA}.
Chang et al. \cite{Chang2007-ServiceAdaptation} present a survey of common adaptation methods in service-oriented computing.
Hirschfeld and Kawamura \cite{Hirschfeld2006-DynamicAdaptation} define adaptability in three dimensions: \emph{what} (e.g., computation/behavior or communication), \emph{when} (e.g., at compile time or run time), and \emph{how} (e.g., composition or transformation).
\par
From a software engineering point of view, variants of services can be realized using \emph{Software Product Lines} (SPL).
SPLs are a development approach for re-usable and interchangeable software \cite{McGregor2002-SPL}.
SPLs are characterized by their variability \cite{VanGurp2001-SPLVariation} and this variability can for instance be represented with \emph{feature models} \cite{Beuche2007-SPLFM}.
Based on such feature models, Sanchez et al. \cite{Sanchez2013-Featuremodels} present a heuristic-based method for the selection of an optimal configuration.
Dynamic software product lines are capable to adapt, e.g., to user requirements or resource constraints \cite{Hallsteinsen2008-DSPL}.
As an example, Weckesser et al. \cite{Weckesser2018-SPL} examine the reconfiguration of dynamic software product lines. Reconfiguration is done based on consistency properties and learned performance-influence models. The authors however do not consider service chains, and hence, cannot capture the interdependencies of adapting multiple services in a service chain.
\par
More recent works have proposed adaptations for microservices in the context of the IoT.
Gholami et al. \cite{Gholami2019-DockerMV} propose the usage of different versions of a microservice (lightweight or heavyweight), primarily for the purpose of scaling the application.
Kannan et al. \cite{Kannan2019-GrandSLAM} present \emph{GrandSLAM}, a microservice execution framework aimed to maximize throughput and reduce SLA violations.
They do not modify the microservices themselves but instead change the request distributions.
It is worth noting that these techniques can be used in conjunction with our proposed approach.
Mendon{\c{c}}a et al.~\cite{Mendonca2018-Adaptation} discuss the trade-off between generality and reusability in self-adaptive microservices.
Bhattacharya and De \cite{Bhattacharya2017-OffloadingAdaptation} survey adaptation techniques in computation offloading, considering only the degree of concurrency and workload heterogeneity as variations in the applications.
Some works present adaptation models for specific applications, e.g. streaming analytics \cite{Zhang2018-Stream}, or to realize fault tolerance \cite{Zhou2015-IoTSA}.
Others adapt the granularity of the services and not the underlying functionalities \cite{Hassan2016-MicroserviceDesign}.
In contrast, we present a general concept for the adaptation of the internal functioning of microservices.

\subsection{Approximate Computing}
\label{sec:backgroud_relatedwork_approximatecomputing}

Approximate computing trades computation quality with the required effort to perform that computation \cite{Mittal2016-ApproximateSurvey}.
The motivation to use approximate computing stems from the fact that in many problem domains of science and engineering, exact results are not required, but only results that are \emph{good enough}.
Examples can be found in the domain of digital signal processing, multimedia, and data analytics.
Besides algorithmic resilience, users are also tolerant of inaccurate results.
Examples are search results in information retrieval or the quality of images and video streams.
In addition, the usage context might also influence the required computation quality \cite{Machidon2020-ApproxMC}.
\par
At the top level, we can distinguish between hardware and software approaches for approximate computing \cite{Moreau2018-Taxonomy}.
Hardware approaches work by introducing imprecise logic components \cite{Gupta2011-ImpreciseAdders,Ye2013-ApproxAdders} or using techniques like voltage overscaling \cite{Mohapatra2011-VoltageScaling}.
In \EC, we cannot implement approximate computing on a hardware level, given that we opportunistically leverage existing, heterogeneous devices over which we have no direct control.
\par
One example of application-level approximate computing is \emph{Foggy\-Ca\-che}~\cite{Guo2018-FoggyCache}.
The authors propose to reuse computation results across devices, based on the observation that similar contextual properties map to the same or similar outcome.
Perez et al. \cite{Perez2017-Mapreduce} have examined the latency-accuracy trade-off in MapReduce jobs when applying approximate computing.
Chippa et al. \cite{Chippa2013-ACResilience} conduct a study in which they analyze the resilience of different applications to result inaccuracies.
As demonstrated in \cite{Agrawal2016-AC}, different combinations of approximate computing techniques can be combined.
%The authors use loop perforation, reduced precision computation and relaxed synchronization on applications from the domain of digital signal processing, robotics, and machine learning.
Their results suggest that up to 50\,\% in execution time can be saved while producing acceptable results.
Other works have demonstrated the potential impact of approximate computing in different application domains, e.g., iterative methods \cite{Zhang2014-Approxit}, image compression \cite{Almurib2018-Compression}, artificial neural networks \cite{Zhang2015-ANN}, and deep learning \cite{Chen2018-DL}.
\par
Few previous works exist that apply approximate computing to domains that are related to \EC.
Zamari et al. \cite{Zamani2017-Edge} combine approximate computing with \EC in an IoT scenario where sensor data is to be sent to the cloud for analytics.
In-transit edge nodes contribute to the analytics by carrying out intermediate computations.
This is coupled with approximate computing techniques on a software level, such as reducing the number of iterations or skipping certain parameter values.
Wen et al. \cite{Wen2018-ApproxIoT} employ a similar approach.
They present \emph{ApproxIoT}, combining approximate computing  (by using only samples of a raw data stream) with hierarchical processing.
Sch{\"a}fer et al. \cite{Schaefer2016-QoC} introduce several metrics for the \emph{quality of computation} (QoC), for example, speed, precision, reliability, costs, and energy.
They extend their \emph{Tasklet} system \cite{Edinger2017-Dev}---an offloading middleware for distributed computing---to provide execution guarantees \WRT these QoC metrics.
Compared to our adaptations, they do so not by modifying the internal functioning of the computation unit but by controlling their distribution.
\par
In a broader context, Pejovic et al. \cite{Pejovic2018-MAC} outline the challenges for approximate computing on mobile devices with a focus on the users' needs.
Similarly, Machidon et al. \cite{Machidon2020-ApproxMC} have noted that the field of approximate computing for mobile devices still lags behind its counterparts in the desktop and server environment.
Using the example of mobile video decoding, the authors demonstrate how the acceptable quality degradation can vary according to the user's current context.

\section{The concept of adaptable microservices}
\label{sec:approach}

Our approach is based on the concept of microservices, i.e., software components that execute a specific task.
Applications are typically composed of multiple such services.
As a special case, services can form \emph{service chains}, in which the output of one microservice is the input for a subsequent microservice.
In the context of \EC, this development method is especially useful, as it allows for fine-grained offloading decisions.
\par
We propose the dynamic \emph{adaptation} of microservices at runtime.
We use the term \emph{service adaptation} to refer to the \emph{internal functioning} of the microservices.
This definition stems from the observation that a particular functionality can be implemented in different ways, leading to many possible \emph{service variants} between which we can switch at runtime. Triggering a switch can be done via an interface that the service exposes, e.g., to a controller that is responsible for orchestrating the services.
This adaptation is orthogonal to other runtime optimizations that can be made in order to provide certain guarantees, e.g., the scaling or migration of microservices to meet execution time guarantees in view of an increased system load.
Migration strategies, however, introduce a considerable overhead, as program code (and possibly execution environments and state) have to be transferred.
Compared to costly migration strategies, we argue that our approach is a suitable alternative because it allows for quick reconfiguration of instance variants and, therefore, service instances can be kept active for a longer period of time.
\par
Contrary to previous approaches, e.g., in the domain of approximate computing (see \Cref{sec:backgroud_relatedwork_approximatecomputing}), our concept of adaptable microservices combines the following three characteristics: (i)~we adapt the internal functioning of a microservice, i.e., we operate on the application level and adaptations are implemented in the program code of the microservices, (ii)~we propose adaptations in three general dimensions, and (iii)~we envision a control entity that automatically selects and changes the service variants at runtime.
\par
We make microservices adaptable in the following three dimensions:
\begin{enumerate}
  \item \textbf{Algorithms:} A task can typically be performed by a variety of algorithms.
  Those not only differ in their runtime complexity, and hence, result in varying execution time, hardware requirements, and energy consumption, but also in their suitability for different applications.
  Taking the example of compressing an image, some compression algorithms are better suited for photographs while others perform better on vector graphics.
  \item \textbf{Parameters:} Parameters are variable inputs to the microservice that influence its execution behavior.
  We model parameters as key-value pairs.
  Parameters can, for example, customize the algorithm that is used.
  Taking the same example of image compression, the desired image quality would be a parameter for such a microservice.
  Parameters can also be used to explicitly limit the execution time of a microservice, e.g., via loop perforation\footnote{loop perforation refers to skipping certain iterations in a loop or breaking the loop after a number of iterations.} \cite{Sidiroglou2011-LoopPerforation}.
  \item \textbf{Auxiliary Data:} Some algorithms require auxiliary data to function.
  This data is often retrieved from external sources.
  An example in the domain of machine learning are pre-trained models.
  This auxiliary data can also influence the execution time and the computation result.
  For example, in recognition tasks performed by neural networks, more complex models produce more accurate results, but require more computing resources or take longer to complete the task.
\end{enumerate}

\begin{figure}[!t]
  \centering
  \includegraphics[scale=0.77]{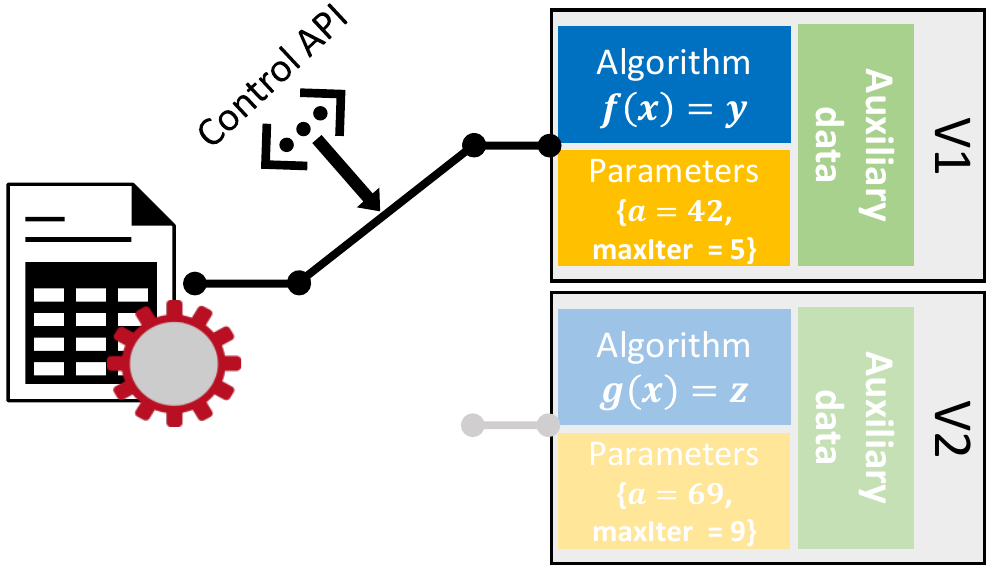}
  \caption[Switching between variants of an adaptable microservice]{Switching between variants of an adaptable microservice}
  \label{fig:SA_servicevariants}
\end{figure}

\Cref{fig:SA_servicevariants} visualizes the concept of adaptable microservices.
A given service might be variable in one or more of the aforementioned dimensions.
We define the possible combinations of all three adaptation dimensions as \emph{service variants}. As an example, the service in \Cref{fig:SA_servicevariants} has two different variants.

\begin{definition}
Given a set of implemented algorithms $\mathcal{A}$, parameters $\mathcal{P}$, and auxiliary data $\mathcal{D}$ for a microservice, a \textbf{service variant} $Var_M$ of a microservice $M$ is defined as $Var_M \subseteq \mathcal{A} \times \mathcal{P} \times \mathcal{D}$ with $p_i = v_i, i=1 \ldots n$ as values for the parameters.
Note that $\mathcal{A}$ and $\mathcal{D}$ are finite sets, whereas $\mathcal{P}$ typically is an uncountable set, e.g., in case the parameters contain real numbers.
\end{definition}
% Streichkandidat
We assume that there are no variants across a service chain that are mutually exclusive. Should one want to consider this case, constraint solvers can be used for selecting valid variants \cite{Benavides2010-Featuremodels}.
We further assume that each of the variants is implemented in the microservice.
For example, if a microservice can be implemented using different algorithms, all those algorithms are included in the source code of the service.
At any given time, a service maps its current variant to an internal state that determines how it is executed when requests are processed.
\par
The different service variants impact the result of the computation in two ways.
First, the execution time varies, e.g., when less complex algorithms are invoked or loop iterations are skipped.
Naturally, this leads to a reduction in energy consumption of the cloudlet which executes the microservice.
Second, service variants impact the quality of result (QoR).
Depending on the application, QoR needs to be defined differently.
We can divide QoR-metrics into two categories: (i)~user-centered and (ii)~numeric.
For user-centered metrics, techniques like questionnaires or focus groups can be used to assess the perceived quality of result.
Note that this might not only vary from one user to another but also might depend on the usage context (as noted in \cite{Machidon2020-ApproxMC}).
As a numeric metric, we can for example quantify the error in the computation, i.e., the deviation from a numeric optimum or the accuracy of the result.

\section{Realization of the concept at the edge}
\label{sec:implementation}

\subsection{Integration into an \EC Framework}
\label{subsec:implementation_framework}

\begin{figure}[!htpb]
\centering
\includegraphics[scale=0.8]{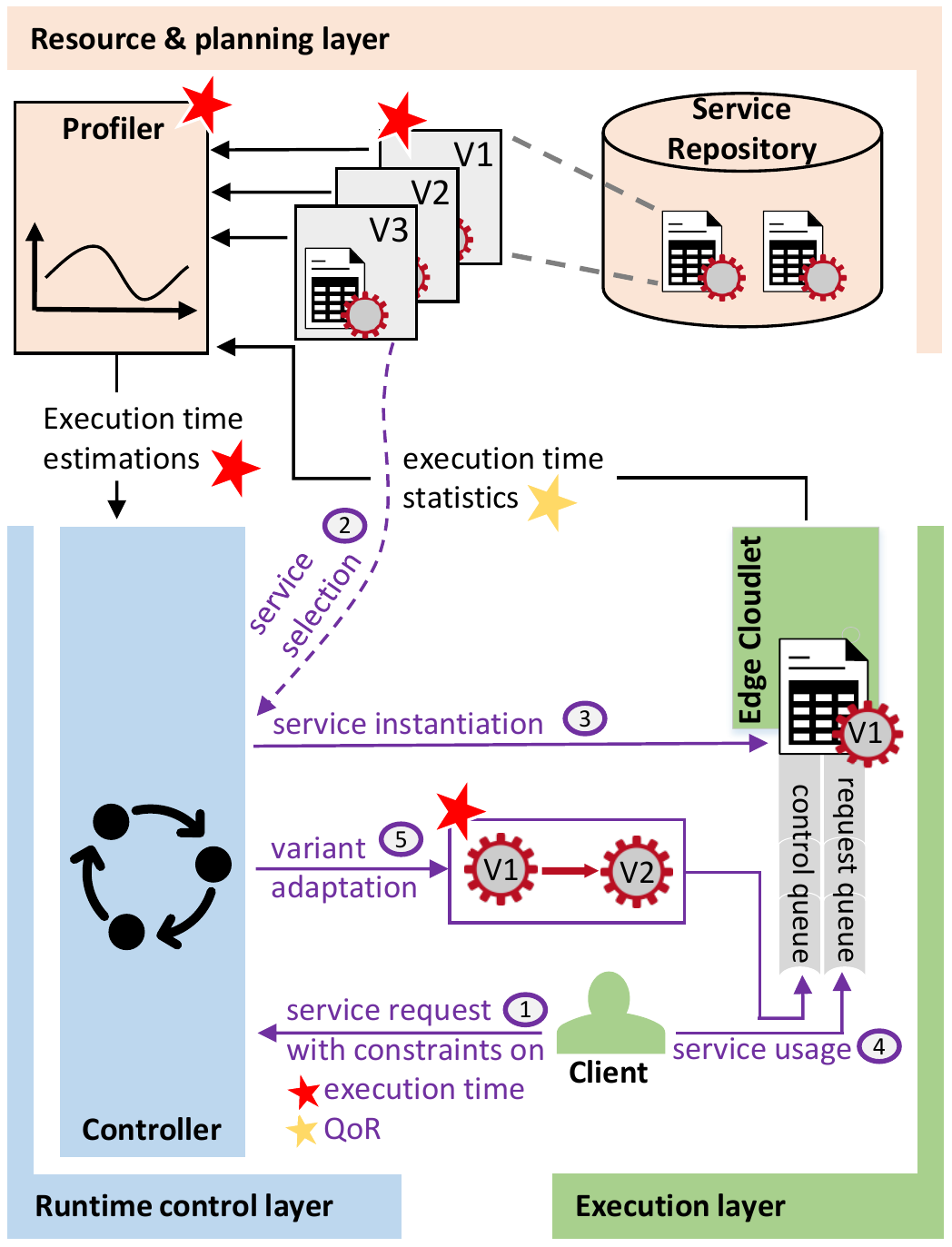}
\caption[Adaptable microservices in an \EC framework]{Adaptable microservices in an \EC framework}
\label{fig:implementation_framework}
\end{figure}

To realize our concept, we implement the required mechanisms for adaptable microservices into our previously developed \EC framework \cite{gedeon2019-microserviceOffloading}.
\Cref{fig:implementation_framework} depicts an overview of this design.
Components that are specifically related to the adaptable microservices are marked with a red star in \Cref{fig:implementation_framework}.
Parts of the implementation that we leave open for future work are marked with a yellow star.
\par
We structure the design of our system into three layers: (i)~a resource and planning layer that provides the adaptable microservices and profiling of the services, (ii)~the runtime control layer that manages the variants of the services, and (iii)~the execution layer, where the service variants run on the edge agents.
In our implementation, microservices are encapsulated as Docker containers that run on edge cloudlets.
The microservices are made available through a repository.
Users address requests for service execution to a \emph{controller} that is responsible for orchestrating the services.
Communication with the microservices is done via asynchronous message queues.
Our implementation uses \emph{RabbitMQ} as a message broker.
%\subsubsection{Including adaptive microservices}
%\label{subsubsec:implementation_framework_adaptations}

\paragraph*{Profiling and monitoring of adaptive services}
Variants of a microservice are included in its implementation.
For each variant, an offline profiler creates a model to estimate the execution time, given different input sizes and underlying hardware on which the service is executed.
\Cref{subsec:evaluation_runtime} will demonstrate the accuracy of such a profiling.
This information serves as a basis for the controller for choosing suitable variants at the start of a microservice or changing the variants of running services.
Given the heterogeneity of execution environments, not all possible hardware configurations can be considered.
Hence, this information should be gradually updated with collected runtime statistics from the agents.

\paragraph*{Control flow}
Users can submit their requests for the execution of a service with a constraint on the execution time or the quality of result (shown as \ding{172} in \Cref{fig:implementation_framework}).
Based on these constraints and the information from the profiler, a suitable service variant is selected  \ding{173}, instantiated \ding{174}, and can then serve user requests \ding{175}.
Note that for simplicity reasons, the figure only depicts a single microservice.
For service chains, the decision-making process is made across all services in the chain.
The question of how to place and spread the placement of individual microservices belonging to a service chain is beyond the scope of our paper.
Previous works (e.g., \cite{gedeon2018-operatorPlacement}) have investigated this problem to great length.

\paragraph*{Changing service variants}
During the execution of a microservice, its variant can be changed.
This is done by issuing requests to a dedicated \emph{control queue} for each microservice instance.
As an example, in \Cref{fig:implementation_framework}, the service variant is changed from \emph{V1} to \emph{V2} \ding{176}.
Microservices implement a listener for incoming requests on the control queue and change their variant accordingly.
This adaptation at runtime can be done for a number of reasons, e.g., when a constraint on the execution time cannot be met, a service might be instructed by the controller to switch to a variant that produces less accurate but faster results.
\Cref{subsec:evaluation_switching} will explore the practical impact of variant switching.

\subsection{Demo Microservices}
\label{subsec:implementation_microservices}

We demonstrate our approach by using six individual microservices and two service chains.

\begin{table*}[!thbp]
  \small
  \centering
  \caption[Summary of service variants]{Summary of service variants}
  \label{tbl:implementation_servicevariants}
  \begin{tabular}{l|ccc}
  \toprule
  \multirow{2}{*}{\textbf{Microservice}} & \multicolumn{3}{c}{\textbf{Variants}} \\
                & Algorithms    & Parameters    & Auxiliary Data    \\
  \midrule
  Face detection& $\{$\textit{LBP-Classifier},\textit{Haar-Classifier}$\}$  &  $\{$\textit{scale-factor},\textit{min-neighbors} $\}$   & $\emptyset$  \\
  \rowcolor{tablegray}
   & & & $\{$\textit{faster\_rcnn\_inception\_v2\_} \\
  \rowcolor{tablegray}
  & & & \textit{coco, ssd\_mobilenet\_v1\_coco,} \\
  \rowcolor{tablegray}
  & & & \textit{ssd\_mobilenet\_v1\_fpn}, \\
  \rowcolor{tablegray}
  \multirow{-2}{*}{Object} & & & \textit{ssd\_mobilenet\_v1\_ppn}, \\
  \rowcolor{tablegray}
  \multirow{-2}{*}{detection} & & & \textit{ssd\_mobilenet\_v2\_coco,} \\
  \rowcolor{tablegray}
  & & & \textit{ssd\_resnet50\_v1\_fpn,} \\
  \rowcolor{tablegray}
  & \multirow{-7}{*}{$\emptyset$} &   \multirow{-7}{*}{$\emptyset$} & \textit{ssdlite\_mobilenet\_v2\_coco}$\}$ \\
  Image & & $\{$\textit{compression-} & \\
  compression  &   \multirow{-2}{*}{$\emptyset$} & \textit{quality}$\}$ &   \multirow{-2}{*}{$\emptyset$} \\
  \rowcolor{tablegray}
  Image &  $\{$\textit{Gaussian blur}, & & \\
  \rowcolor{tablegray}
  blurring & \textit{Median blur}$\}$  &   \multirow{-2}{*}{$\{$\textit{kernel-size}$\}$}  & \multirow{-2}{*}{$\emptyset$} \\
  Image & & & $\{$\textit{psnr-large, psnr-small,}, \\
  upscaling & \multirow{-2}{*}{$\emptyset$} & \multirow{-2}{*}{$\emptyset$} & \textit{noise-cancel, gans}$\}$ \\
  \rowcolor{tablegray}
  3D mesh re-& & & $\{$\textit{meshrcnn, pixel2mesh},\\
  \rowcolor{tablegray}
  construction & \multirow{-2}{*}{$\emptyset$} & \multirow{-
  2}{*}{$\emptyset$} & \textit{sphereinit, voxelrcnn}$\}$ \\
  \bottomrule
  \end{tabular}
\end{table*}

\subsubsection{Individual microservices}
\label{subsubsec:implementation_microservices_single}

We implement the following adaptable microservices, summarized with their different variants in \Cref{tbl:implementation_servicevariants}: \\
  \bpara{Face detection:} This microservice detects faces in a given picture. Its variants differ in algorithms and parameters.
  For the algorithms, we use two different types of cascade classifiers available in OpenCV: (i)~\emph{LBP} and (ii)~\emph{Haar}.
  In general, LBP is faster but produces less accurate results.
  The microservice furthermore expects two parameters: (i)~\emph{scale-factor} and (ii)~\emph{min-neighbors}.
  The first parameter determines the scaling between two levels of upscaling or downscaling (because both algorithms work only on predefined model dimensions).
  The second parameter \emph{min-neighbors} specifies the minimum number of neighbors for candidate rectangles for those to be retained.
  Higher values for this parameter lead to fewer faces being detected but at the same time, this also decreases the number of false positives. \\
  \bpara{Object detection:} This microservice uses \emph{TensorFlow} to detect objects in a given image.
	The microservice uses different auxiliary data with pre-trained models\footnote{\url{https://github.com/tensorflow/models/blob/master/research/object\_detection/g3doc/detection\_model\_zoo.md} (accessed: 2020-04-22)}.
	The models differ in their execution speed and mean average precision.
  \\
  \bpara{Image compression:} Using the image encoding function of OpenCV, this microservice compresses a given input image using JPEG.
  As the only variation, the compression quality can be specified as a parameter. \\
  \bpara{Image blurring:} Given an input image and an array of rectangular regions, this microservice blurs the given regions of the image.
  To perform the operation, we use OpenCV's blur function.
  The blurring can be performed by two different algorithms: (i)~\emph{Gaussian blur} and (ii)~\emph{median blur}.
  The Gaussian blur is a linear filter that is faster but does not preserve edges in the original image.
  In contrast, the median blur is a non-linear filter that is able to preserve edges.
  For both algorithms, a \emph{kernel size} is used as a parameter to determine the size of the convolution matrix. \\
  \bpara{Image upscaling:} This microservice produces an upscaled image of the input image.
  It also aims at enhancing the quality of the upscaled image by using Residual Dense Networks (RDN).
  We use an existing Keras\footnote{\url{https://keras.io/} (accessed: 2020-04-21)}-based implementation\footnote{\url{https://github.com/idealo/image-super-resolution} (accessed: 2020-04-16)} as a basis for our microservice.
  We use four different pre-trained models that are variants of auxiliary data: \emph{psnr-large}, \emph{psnr-small}, \emph{noise-cancel}, and \emph{gans}.
  Except for the \emph{gans} model (which quadruples the resolution), these models double the original image resolution.\\
  \bpara{3D mesh reconstruction:} This microservice aims at reconstructing a 3D mesh representation of an object in a (2D) picture.
  We use the published code\footnote{\url{https://github.com/facebookresearch/meshrcnn} (accessed: 2020-04-16)} of Gkioxari et al. \cite{Gkioxari2019-3DMesh} as the basis for our microservice.
  Four different models are used as auxiliary data.
  Some reconstruct only the shape of the object while others use voxels to achieve a more realistic representation of the object.

  \begin{figure*}[!htbp]
  \centering
  \subfigure[Face anonymization]{\label{fig:implementation_pipelines_faceblur}\includegraphics[scale=0.41]{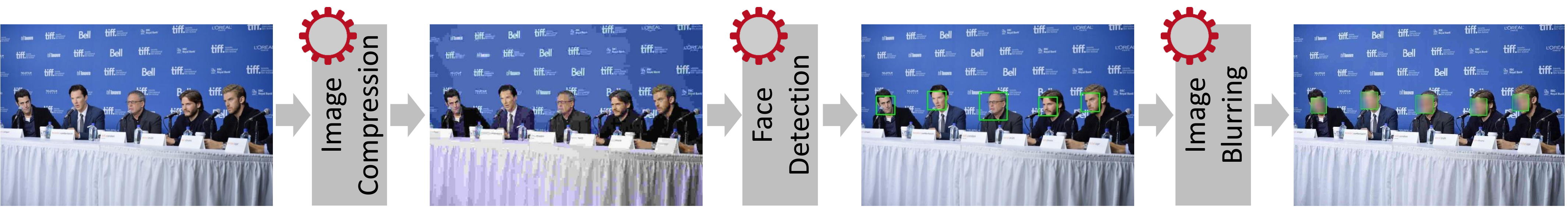}}
  \subfigure[3D mesh reconstruction of upscaled images]{\label{fig:implementation_pipelines_3dmesh}\includegraphics[scale=0.46]{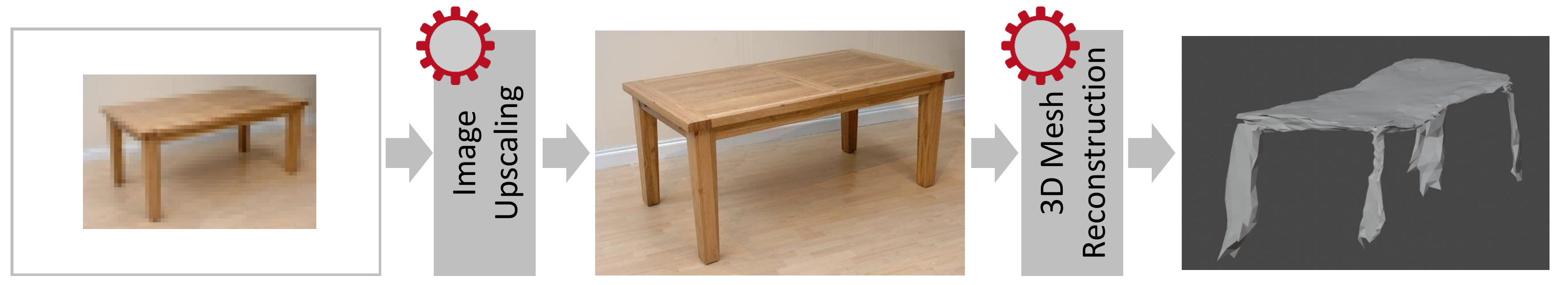}}
  \label{fig:implementation_eval_pipelines}
  \caption[Microservice chains]{Microservice chains}
  \end{figure*}

\subsubsection{Microservices chains}
\label{subsubsec:implementation_microservices_chains}

From the individual microservices we construct two service chains.\\
\bpara{Face anonymization:} Given an image as input, this chain anonymizes faces by blurring them. First, the original image is compressed.
Afterward, a face detection is performed, outputting detected faces as rectangular coordinates to the next service.
As a final step, the image blurring microservice blurs the regions returned by the face detection microservice.
An illustrative example of this service chain is shown in \Cref{fig:implementation_pipelines_faceblur}. \\
\bpara{3D mesh reconstruction of upscaled images:} This microservice chain first performs an upscaling of an input image and then reconstructs a 3D mesh from the upscaled image.
\Cref{fig:implementation_pipelines_3dmesh} illustrates an example execution of this chain.

\section{Orchestrating microservice variants}
\label{sec:evaluation}

In this section, we study the practical impact of microservice variants.
First, we study how accurately we can model the execution time of the microservices and which features are relevant for creating a model of the execution time that is as accurate as possible (\Cref{subsec:evaluation_runtime}).
We then study how different variables of service variants correlate and what their impact on the execution time and quality of result is (\Cref{subsec:evaluation_variants}).
Lastly, we demonstrate how switching of service variants at runtime can help adapt to varying workloads in view of execution time constraints (\Cref{subsec:evaluation_switching}).

\subsection{Execution Time Estimation}
\label{subsec:evaluation_runtime}

\begin{table}[!ht]
\small
\centering
\caption[AWS Instance types used for benchmarking]{AWS Instance types used for benchmarking}
\label{tbl:evaluation_ec2machines}
\begin{tabular}{lccc}
\toprule
\cellcolor{white}\textbf{Type} & \textbf{vCPUs} & \textbf{Clock rate} & \textbf{Memory} \\ \midrule
t2.micro & 1 & \unit[2.5]{GHz}$^{a}$ & \unit[1]{GB} \\
\rowcolor{tablegray}
t2.small & 1 & \unit[2.5]{GHz}$^{a}$ & \unit[2]{GB} \\
t2.medium & 2 & \unit[2.3]{GHz}$^{b}$ & \unit[4]{GB}  \\
\rowcolor{tablegray}
t2.large & 2 & \unit[2.3]{GHz}$^{b}$ & \unit[8]{GB}  \\
t2.xlarge & 4 & \unit[2.3]{GHz}$^{b}$ & \unit[16]{GB} \\
\rowcolor{tablegray}
t2.2xlarge & 8 & \unit[2.3]{GHz}$^{b}$ & \unit[32]{GB} \\
c5.large & 2 & \unit[3.4]{GHz}$^{c}$ & \unit[4]{GB} \\
\rowcolor{tablegray}
c5.xlarge & 4 & \unit[3.4]{GHz}$^{c}$ & \unit[8]{GB} \\
c5.4xlarge & 16 & \unit[3.4]{GHz}$^{c}$ & \unit[32]{GB}  \\
\rowcolor{tablegray}
r5d.large & 2 & \unit[3.1]{GHz}$^{d}$ & \unit[16]{GB} \\
r5d.2xlarge & 8 & \unit[3.1]{GHz}$^{d}$ & \unit[64]{GB} \\
\bottomrule
\multicolumn{4}{l}{$^{a}$Intel Xeon Family} \\
\multicolumn{4}{l}{$^{b}$Intel Broadwell E5-2695v4} \\
\multicolumn{4}{l}{$^{c}$Intel Xeon Platinum 8124M} \\
\multicolumn{4}{l}{$^{d}$Intel Xeon Platinum 8175} \\
\end{tabular}
\end{table}

\paragraph*{Methodology and experimental setup} To estimate the execution time of microservices, we build regression models using supervised learning methods.
We use three variants of estimators implemented in \emph{scikit-learn}\footnote{\url{https://scikit-learn.org} (accessed: 2020-04-13)}, a machine learning library for Python: (i)~decision trees regressor, (ii)~random forest regressor, and (iii)~extra tree regressor.
For each estimated model, the \emph{R\textsuperscript{2}-score} is computed to assess its quality.
This metric gives an indication of how accurate the model is.
To analyze the impact of different underlying hardware configurations, we run this evaluation on different AWS EC2 instance types as summarized in \Cref{tbl:evaluation_ec2machines}.
They differ in CPU and memory configuration and are optimized for either general-purpose (t-type instances), computing (c-type instances), or memory-intensive (r-type instances) applications.
For each instance type, we run benchmarks of the face detection and object detection microservice.
During the execution of the microservices, we also record statistics on available hardware resources and system load.
Those will serve as possible features to build a model of the execution time (for instance, to be able to predict the execution time given different load levels on the system).
For face detection, we use the two different face detection algorithms, \emph{LBP} and \emph{Haar}~\cite{Kadir2014-FaceDetectionClassifiers}.
The \emph{scale-factor} parameter is varied from 1.0 to 1.9 in 0.1 increments and values for \emph{min-neighbors} are varied from 1 to 10.
We use 46\,160 different images from the \emph{WIDER FACE}\footnote{\url{http://shuoyang1213.me/WIDERFACE/} (accessed: 2020-04-25)} dataset.
For the object detection, we use two different models (\emph{ssd\_mobilenet\_v1\_coco} and \emph{faster\_rcnn\_inception\_v2\_coco}) on the \emph{val2017} dataset included in the \emph{Coco Dateset}\footnote{\url{http://cocodataset.org/} (accessed: 2020-04-25)}.
For each instance type, we list the combination of features and regression methods that lead to the highest $R^2$-score.
Features can be properties related to the variant of the microservice (e.g., a parameter) or attributes of the machine where it is executed.
\Cref{tbl:evaluation_estimation_facedetection} shows the results for the face detection and \Cref{tbl:evaluation_estimation_objectdetection} the results for the object detection.
In the tables, the features are ordered in decreasing order of importance, i.e., to what extent they contribute to the prediction of the execution time.

\begin{table}[!ht]
\small
\centering
\caption[Execution time estimation results for the face detection microservice]{Execution time estimation results for the face detection microservice}
\label{tbl:evaluation_estimation_facedetection}
\begin{tabular}{llc}
\toprule
\textbf{Instance} & \textbf{Features}$^{a}$ & \textbf{$R^2$-Score} \\
\midrule
t2.micro & SF, DF-A, MN & 0.4675 \\
\rowcolor{tablegray}
t2.small & CF, DF-A, MEM-A & 0.4317  \\
t2.medium & CF, CPU-U, DF-C  & 0.8717 \\
\rowcolor{tablegray}
t2.large & CPU-U, CF, SF, NF & 0.9456 \\
t2.xlarge & CPU-U, CF, SF, MN, NF & 0.9842 \\
\rowcolor{tablegray}
t2.2xlarge & CF, SF, CPU-U, NF & 0.9856 \\
c5.large & CF, SF, CPU-U, NF  & 0.9887 \\
\rowcolor{tablegray}
c5.xlarge & CF, SF, CPU-U, NF & 0.9914 \\
\multirow{2}{*}{c5.4xlarge} & CF, CPU-U, SF, CPU-F, & \multirow{2}{*}{0.9965} \\
& DF-C, MEM-A, MEM-U & \\
\rowcolor{tablegray}
r5d.large & CF, SF, CPU-U, NF & 0.9883 \\
r5d.2xlarge & CF, SF, CPU-U, NF & 0.9860 \\
\bottomrule
\multicolumn{3}{l}{$^{a}$\footnotesize{\textbf{CF:} classifier, \textbf{CPU-F:} CPU frequency, \textbf{CPU-U:} CPU usage,}} \\
\multicolumn{3}{l}{\footnotesize{~~\textbf{DF-A:} detected faces (absolute number),}} \\
\multicolumn{3}{l}{\footnotesize{~~\textbf{DF-C:} detected faces (correct percentage),}} \\
\multicolumn{3}{l}{\footnotesize{~~\textbf{MEM-A:} available memory, \textbf{MEM-U:} used memory,}} \\
\multicolumn{3}{l}{\footnotesize{~~\textbf{MN:} min-neighbors, \textbf{NF:} number of faces, \textbf{SF:} scale-factor}} \\
\end{tabular}
\end{table}

\begin{table}[!ht]
\small
\centering
\caption[Execution time estimation results for the object detection microservice]{Execution time estimation results for the object detection microservice}
\label{tbl:evaluation_estimation_objectdetection}
\begin{tabular}{llc}
\toprule
\textbf{Instance} & \textbf{Features}$^{a}$ & \textbf{$R^2$-Score} \\
\midrule
t2.micro &  \multicolumn{2}{c}{n.a$^{b}$}  \\
\rowcolor{tablegray}
t2.small & M, CPU-U, MEM-T & 0.5651 \\
\multirow{2}{*}{t2.medium} & MEM-AP, M, MEM-U,  & \multirow{2}{*}{0.6938}  \\
& C, CPU-F & \\
\rowcolor{tablegray}
t2.large & M, MEM-A, MEM-U & 0.6827 \\
t2.xlarge & M, MEM-A, MEM-T & 0.6412 \\
\rowcolor{tablegray}
t2.2xlarge & M, MEM-AP, CPU-U & 0.7690  \\
c5.large & M, CPU-U, MEM-T & 0.9970 \\
\rowcolor{tablegray}
c5.xlarge & M, CPU-U, MEM-T & 0.9969 \\
c5.4xlarge & M, CPU-U, MEM-T & 0.9968 \\
\rowcolor{tablegray}
r5d.large & M, CPU-U, MEM-T & 0.9968 \\
r5d.2xlarge & M, CPU-U, MEM-T  & 0.9970 \\
\bottomrule
\multicolumn{3}{l}{$^{a}$\footnotesize{\textbf{C:} correctness, \textbf{CPU-F:} CPU frequency, \textbf{CPU-U:} CPU usage,}}  \\
\multicolumn{3}{l}{\footnotesize{~~\textbf{MEM-AP:} available memory (percentage),}} \\
\multicolumn{3}{l}{\footnotesize{~~\textbf{MEM-A:} available memory (absolute), \textbf{M:} model,}} \\
\multicolumn{3}{l}{\footnotesize{~~\textbf{MEM-T:} total memory, \textbf{MEM-U:} used memory}} \\
\multicolumn{3}{l}{$^{b}$\footnotesize{hardware configuration not sufficient to run the microservice}} \\
\end{tabular}
\end{table}

\paragraph*{Impact of the machine types} From the results, we can observe that especially with the more powerful machines, we can achieve high $R^2$-scores and, hence, a high accuracy of the model.
For less powerful types of machines, e.g., \emph{t2.micro} and \emph{t2.small}, we get much lower scores.
This is likely due to a greater variance in execution times that happens because \emph{t2}-type instances are so-called \emph{burstable instances}, i.e., if the system is overloaded, the CPU performance of the virtual machine is temporarily increased.
Since this is likely to happen with the least powerful types we used, the high variance of execution times is due to the constant on-off switching of the performance boost.

\paragraph*{Differences in estimators and features} As a second observation, in all but one case (face detection on a \emph{c5.4xlarge} instance), the \emph{extra tree} regressor led to the highest $R^2$-score.
We can also observe great differences in the most relevant features for the execution time estimation.
These differences can both be seen within one microservice, depending on the instance type, and across microservices.
The estimation for the face detection mostly used the classification algorithm as the most relevant feature.
With more powerful hardware, the classifier, the \emph{scale-factor} parameter, and the current CPU usage are consistently ranked the most relevant features, while for less powerful machines, the \emph{min-neighbors} parameter and the number of detected faces were included in the features.
\par
Contrary to the face detection microservices, for the object detection, we can see a clearer division of relevant features depending on the instance type.
While for \emph{t2}-type instances, the available memory is always a highly ranked feature (except for the \emph{t2.small} instance), this changes in favor of the CPU utilization for \emph{c}-type and \emph{r}-type machines.
Another difference is that the \emph{t2}-type instances lead to significantly lower accuracies of the model, as shown by the $R^2$-score.
\paragraph*{Summary} In summary, this analysis of execution time estimators has shown that we are able to accurately profile the different variants of microservice.
In an \EC framework where adaptable microservices are integrated (see \Cref{subsec:implementation_framework}), this step would be performed offline and serve as base knowledge for runtime decisions.
However, we could also observe that this estimation has to be tuned to the individual microservice \WRT the selection of the hardware and features that are used for the estimation.

\subsection{Impact of Service Variants}
\label{subsec:evaluation_variants}

\paragraph*{Methodology} We measure the correlation between different variables that relate to the service variants and the outcomes of the computations.
Most importantly, we want to assess the change in execution time.
In addition, for the face detection algorithm and, consequently, for the face anonymization service chain, we also analyze the impact on the quality of the result.
\par
To measure the pairwise correlation between variables, we use the \emph{Kendall rank correlation coefficient} throughout this section.
Contrary to other metrics for correlation, such as the \emph{Pearson correlation coefficient}, it has the advantage that it does not assume a linear relationship between variables.

\begin{figure}[!htbp]
  \centering
  \includegraphics[scale=0.49]{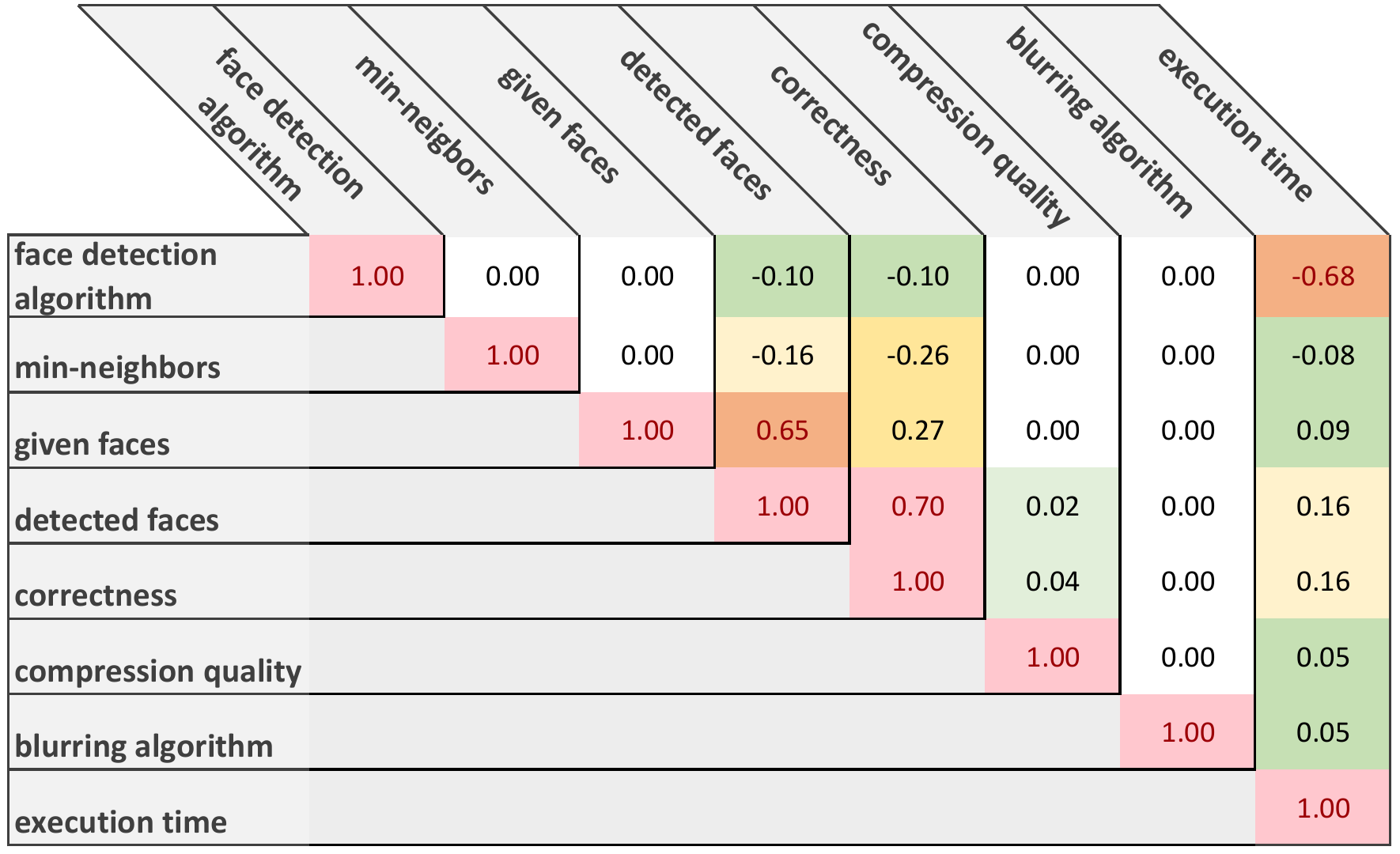}
  \caption[Face blurring chain: correlation matrix of variants]{Face blurring chain: correlation matrix of variants}
  \label{fig:evaluation_corr_facepipeline}
\end{figure}

\paragraph*{Face anonymization service chain} For the first chain, we vary the image compression quality from 1--99 (in steps of 1). We use the two face detection algorithms as described before.
The \emph{scale-factor} parameter is set to a constant 1.2, and \emph{min-neighbors} are varied from 0--9 (step size 1).
For the final step, the blurring microservice, we use \emph{gaussian blur} and \emph{median blur} algorithms with a fixed \emph{kernel size} of $(23,23)$.
We select 21 images and manually label the correct positions of the faces.
Hence, with a small degree of tolerance, besides the absolute number of detected faces, we can also compute a \emph{correctness} value that serves as a metric for the QoR.
For each image and combination, we executed the chain five times and averaged the results.
\par
\Cref{fig:evaluation_corr_facepipeline} shows the correlation matrix of the entire chain.
We can observe that the highest correlation value is attained among the face detection algorithm and the execution time.
To map this correlation to concrete numbers, on average, the execution time using the Haar classifier was \unit[0.13]{s}, while for the LBP classifier it averaged to \unit[0.08]{s}.
This means that by changing the variant of the algorithm, we could achieve a reduction in the execution time of \unit[38.46]{\%}.
However, this reduction in execution time comes at the cost of a reduced correctness value, which drops from 0.67 to 0.57 on average (-\unit[14.92]{\%}).
This provides a good example of the trade-off between the computation complexity (represented by the execution time) and the quality of result (represented by the correct recognition of faces).
\par
Compared to the face detection algorithm, other variables related to the variants, i.e., \emph{min-neighbors}, \emph{compression quality}, and \emph{blurring algorithm} correlate with the execution time with values of -0.08, 0.05, and 0.05, respectively.
It is worth noticing that \emph{min-neighbors} has a much more significant  impact on the correctness (with a correlation value of -0.26) than on the execution time.
\begin{figure}[!htbp]
  \centering
  \subfigure[Image compression]{\label{fig:evaluation_corr_compression}\includegraphics[scale=0.6]{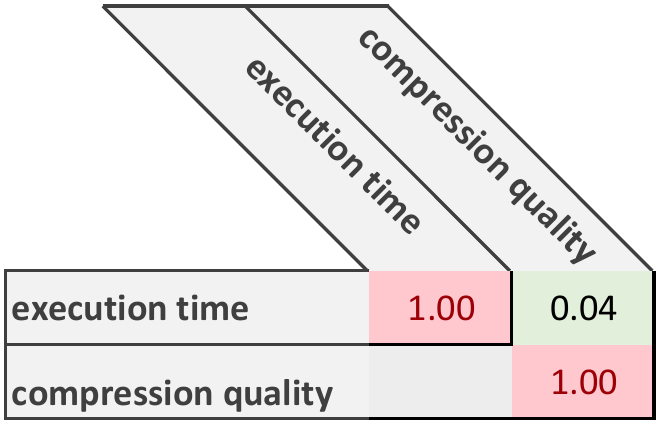}}
  \subfigure[Image blurring]{\label{fig:evaluation_corr_blurring}\includegraphics[scale=0.6]{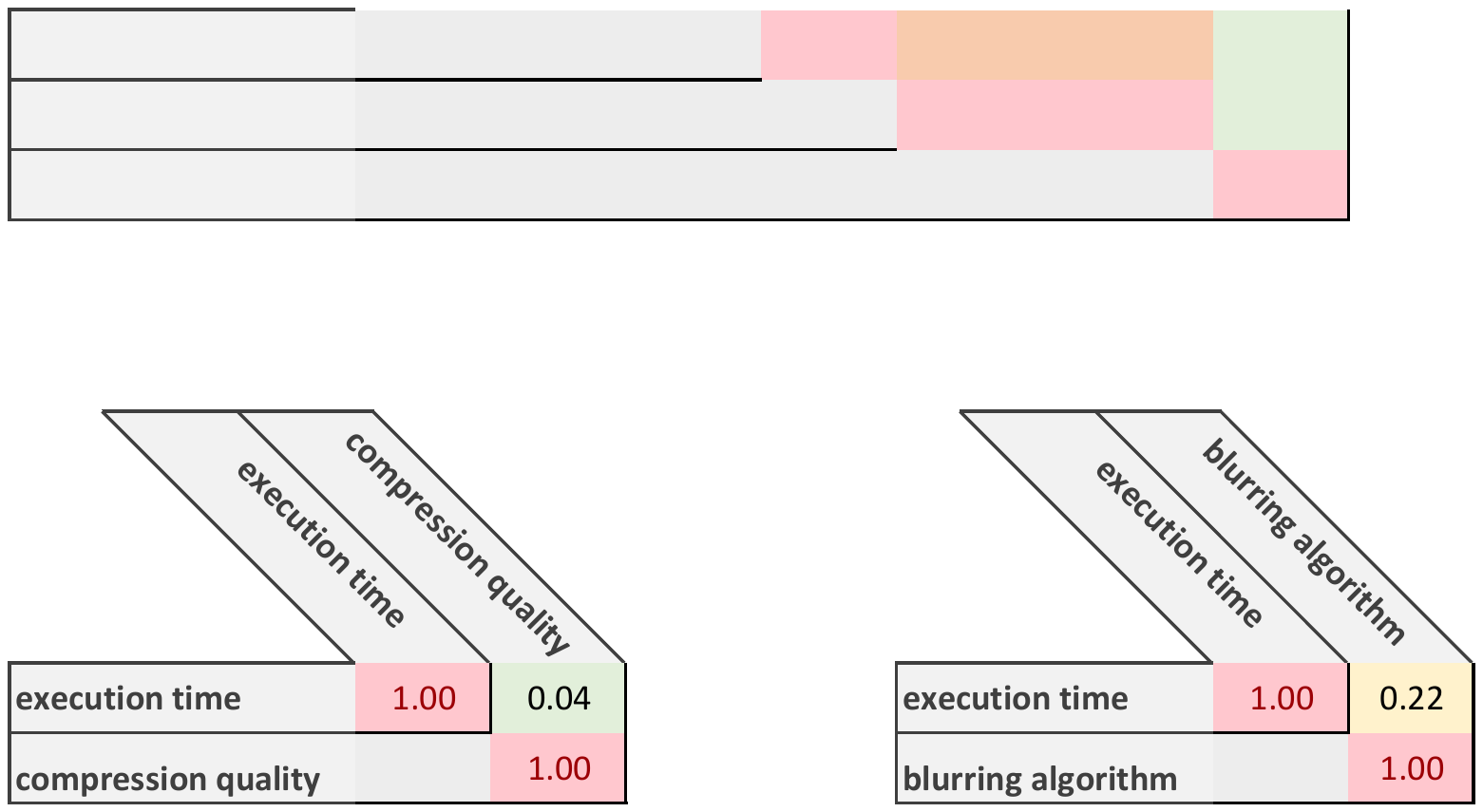}}
  \subfigure[Face detection]{\label{fig:evaluation_corr_facedetection}\includegraphics[scale=0.6]{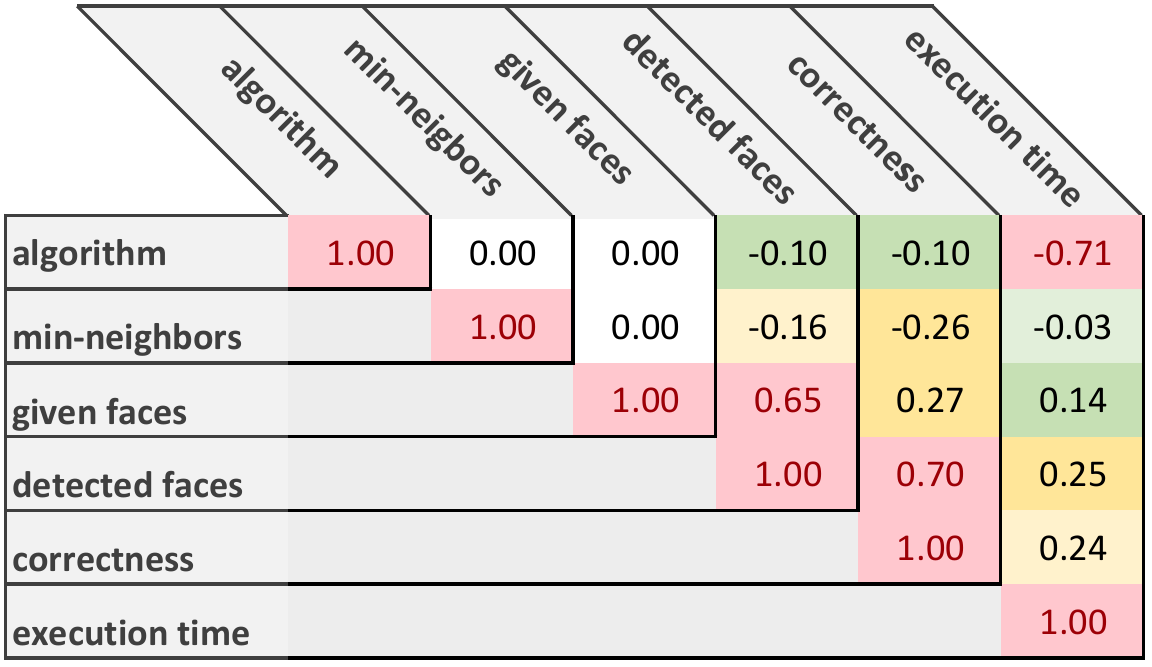}}
  \caption[Correlation matrices for the individual service variants of the face anonymization chain]{Correlation matrices for the individual service variants of the face anonymization chain}
  \label{fig:evaluation_corr_facepipeline_microservices}
\end{figure}
\par
We also provide the correlation matrices of the individual services of this chain in \Cref{fig:evaluation_corr_facepipeline_microservices}.
Comparing \Cref{fig:evaluation_corr_facepipeline_microservices} with \Cref{fig:evaluation_corr_facepipeline} demonstrates the difference in correlation of a single microservice versus when this microservice is integrated into a chain.
As an example, when executed alone, the blurring algorithm has a correlation value of 0.22 with the execution time but in the entire chain, this value drops to 0.05.
A similar change in the correlation score can be observed for the face detection algorithm (-0.68 to -0.71).

\paragraph*{3D mesh reconstruction of upscaled images} For both the image upscaling and 3D mesh reconstruction microservice, we use the four different variants of auxiliary data as previously described.
As input data, we use 5 images from a dataset depicting furniture\footnote{\url{https://www.kaggle.com/akkithetechie/furniture-detector/data} (accessed: 2021-01-07)}.
Because the mesh reconstruction microservice offers GPU support, we execute this service chain on an AWS EC2 \emph{p2.xlarge} instance (Xeon E5-2686 v4, \unit[61]{GB} RAM, Nvidia K80 GPU).
\par
\begin{figure}[!htbp]
  \centering
  \includegraphics[scale=0.6]{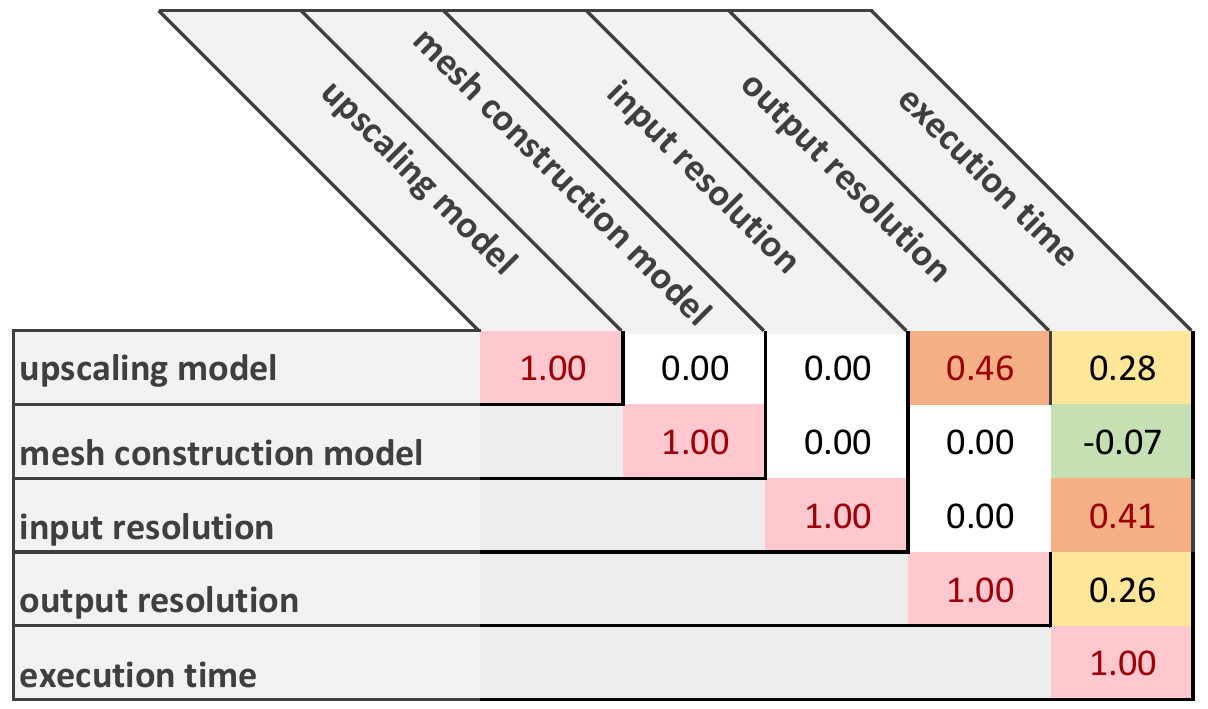}
  \caption[Mesh reconstruction chain: correlation matrix of variants]{Mesh reconstruction chain: correlation matrix of variants}
  \label{fig:evaluation_corr_meshpipeline}
\end{figure}
\begin{figure}[!htbp]
  \centering
  \subfigure[Image upscaling]{\label{fig:evaluation_corr_upscaling}\includegraphics[scale=0.6]{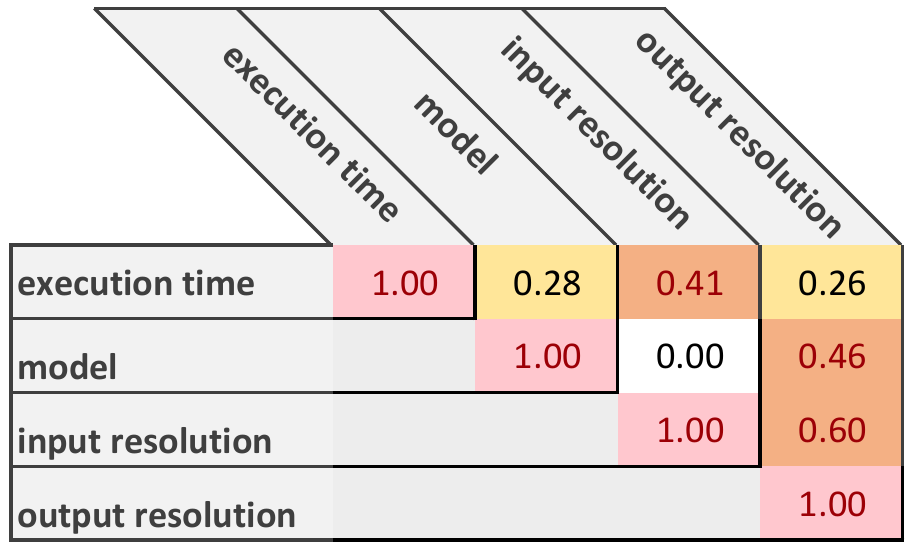}}
  \subfigure[3D mesh reconstruction]{\label{fig:evaluation_corr_meshconstruction}\includegraphics[scale=0.6]{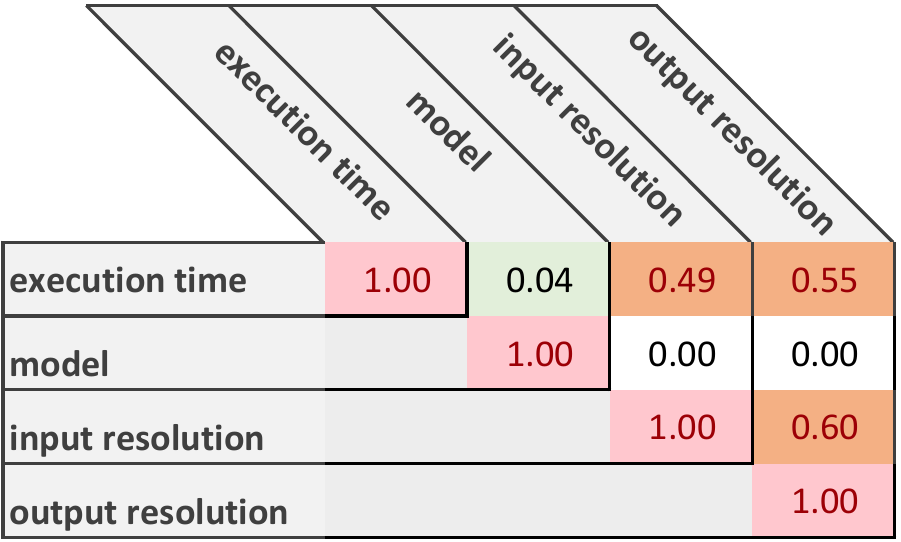}}
  \caption[Correlation matrices for the individual service variants of the mesh reconstruction chain]{Correlation matrices for the individual service variants of the mesh reconstruction chain}
  \label{fig:evaluation_corr_meshpipeline_microservices}
\end{figure}

\Cref{fig:evaluation_corr_meshpipeline} shows the correlation matrix for the entire chain and \Cref{fig:evaluation_corr_meshpipeline_microservices} the matrices for the individual microservices.
Note that for this microservice chain, we leave the exploration of suitable QoR-metrics for future work and focus on the execution times.

\par
The results show that the variants of the upscaling model have more influence than the different mesh reconstruction models (correlation scores of 0.28 and \mbox{-0.07}).
As an example, the \emph{psnr-small} model for image upscaling has an average execution time of \unit[11.55]{s} while the \emph{psnr-large} model averages to \unit[58.94]{s}.
The mean values for the \emph{noise-cancel} and \emph{gans} models are \unit[63.93]{s} and \unit[36.36]{s}, respectively.
This means that by selecting another variant of an image upscaling model, we can reduce the execution time by up to \unit[81.93]{\%}.
In comparison, the differences for the average execution times of the mesh construction models are smaller (\unit[41.26]{s} for \emph{meshrcnn}, \unit[42.01]{s} for \emph{pixel2mesh}, \unit[43.12]{s} for \emph{sphereint}, and \unit[44.38]{s} for \emph{voxelrcnn}).
Hence, here the maximum difference in execution time only amounts to \unit[7.03]{\%}.
Naturally, there is also a strong correlation (0.41 and 0.26) of the execution time with the input and output resolution of the upscaled images.
%When looking at the correlation for the individual microservices (see \Cref{fig:evaluation_corr_meshpipeline_microservices}), these values are even higher (0.41 and 0.60 for the upscaling, and 0.49 and 0.55 for the mesh reconstruction microservice). \TODOJulien{0.60 für upscaling ist der falsche Wert, der ist für die Korrelation zwischen Input und Output resolution (fig 7a). Die richtigen Werte sind 0.41 und 0.26. Damit sind die correlation Werte gleich zwischen Single MS und der Chain beim Upscaling.}

\subsection{Switching of Service Variants}
\label{subsec:evaluation_switching}

\paragraph*{Queue model and variant switching strategy} We model the processing of requests by a microservice as an \emph{M/D/1 queue}.
With this model, we are able to estimate the number of queued messages at which a switch to a faster microservice variant should happen, given a time constraint $C$ that should not be violated.
We assume that requests arrive according to a Poisson process with an arrival rate $\lambda$.
Consequently, the inter-arrival times of requests follow an exponential distribution.
%In \Cref{subsec:evaluation_runtime}, we have shown that the execution time of microservices can be accurately profiled and estimated.
%Based on this observation,
We assume a deterministic service time $D$ in our model, which is the average execution time of a microservice variant on a given hardware setup.
Accordingly, the service rate $\mu$ can be calculated with the formula $\mu = \frac{1}{D}$.

Our strategy for variant switching tries to select the microservice with the highest execution time, so that a given time constraint $C$ is not violated.
This is based on the assumption that more complex microservices deliver better results.
%In general, we assume that microservice variants with higher runtimes deliver better results than variants with lower runtimes.
Furthermore, we assume that the length $L$ of the request queue is known at any given time.
The waiting time $\omega$ of a request is the sum of the time it has to wait in the queue $\omega_{Q}$ and the service time $D$ of the microservice $\omega = \omega_{Q} + D, \omega_{Q} = L * D$.
%\begin{equation}
%	\omega = \omega_{Q} + D, \omega_{Q} = L * D
%\end{equation}
$\omega$ and $C$ are the basis for the estimation of a threshold at which a microservice will switch to a less computing-intensive, and therefore, faster variant.
To avoid a violation of $C$, the queue length $L$ should not exceed a certain threshold $T$, so that the condition $\omega < C$ holds.
$T$ is calculated with $C$ and $D$ as	$T = \lfloor\frac{C}{D} - 1\rfloor$.
To allow for fine adjustments of $T$, we introduce a dampening factor $T_{dampened} = \alpha T, \alpha \in (0,1]$.
A variant switch will be performed if $L > T_{dampened}$.
$T_{dampened}$ is chosen in such a way that violations of $C$ are minimized.
Our variant switching strategy tries to avoid the request queue becoming unstable.
The queue utilization $\rho$ is used to determine the stability of a queue and is given as $\rho = \frac{\lambda}{\mu}$.
If $\rho > 1$, i.e., if $\lambda > \mu$, then the queue becomes unstable because more requests arrive than the microservice can handle and the request queue will fill up.
This will lead to violations of the execution time constraint as more requests in the queue lead to longer waiting times.
To avoid this, our variant switching strategy tries to select a variant that can handle the expected request load, i.e., a variant with a sufficiently high $\mu$ so that $\lambda < \mu$.
Furthermore, the selected variant should not lead to violations of $C$, i.e., for the average wait time of requests $\omega_{avg}$ for the selected variant the condition $\omega_{avg} < C$ should hold.
$\omega_{avg}$ can be calculated as follows:
\begin{equation}
	\label{eq:avg_wait_time}
	\omega_{avg} = \frac{1}{\mu} + \frac{\rho}{2\mu(1-\rho)} = D + \frac{D\rho}{2 (1-\rho)}
\end{equation}

\begin{figure*}[!ht]
	\centering
	\subfigure[Queue length of face detection with $\lambda = 15$]{
		\label{fig:evaluation_variant_switch_face_detection_15_QL_small}
		\includegraphics[height=0.227\textwidth]{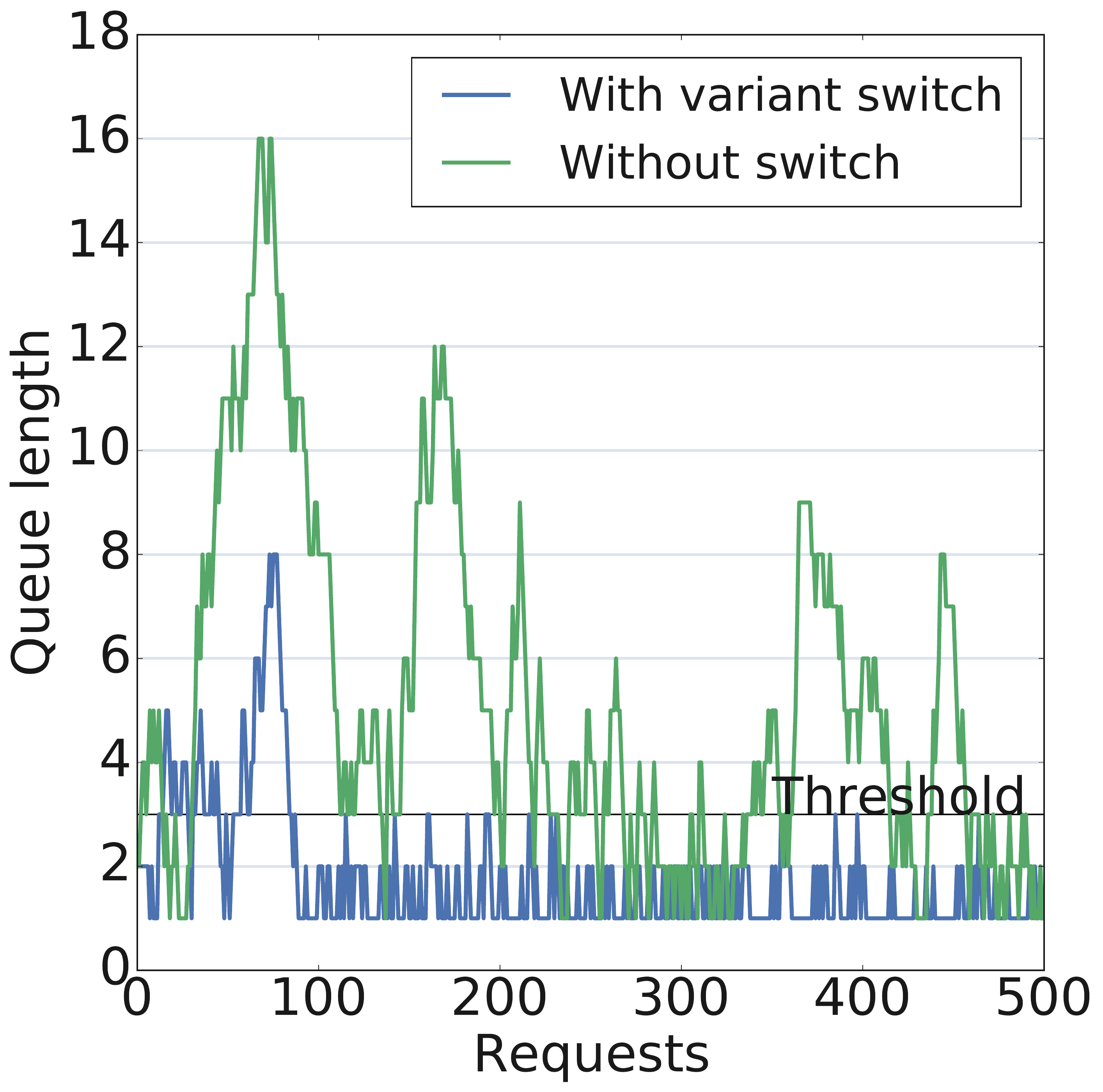}
	}
	\subfigure[Queue length of face detection with $\lambda = 17$]{
		\label{fig:evaluation_variant_switch_face_detection_17_QL_small}
		\includegraphics[height=0.227\textwidth]{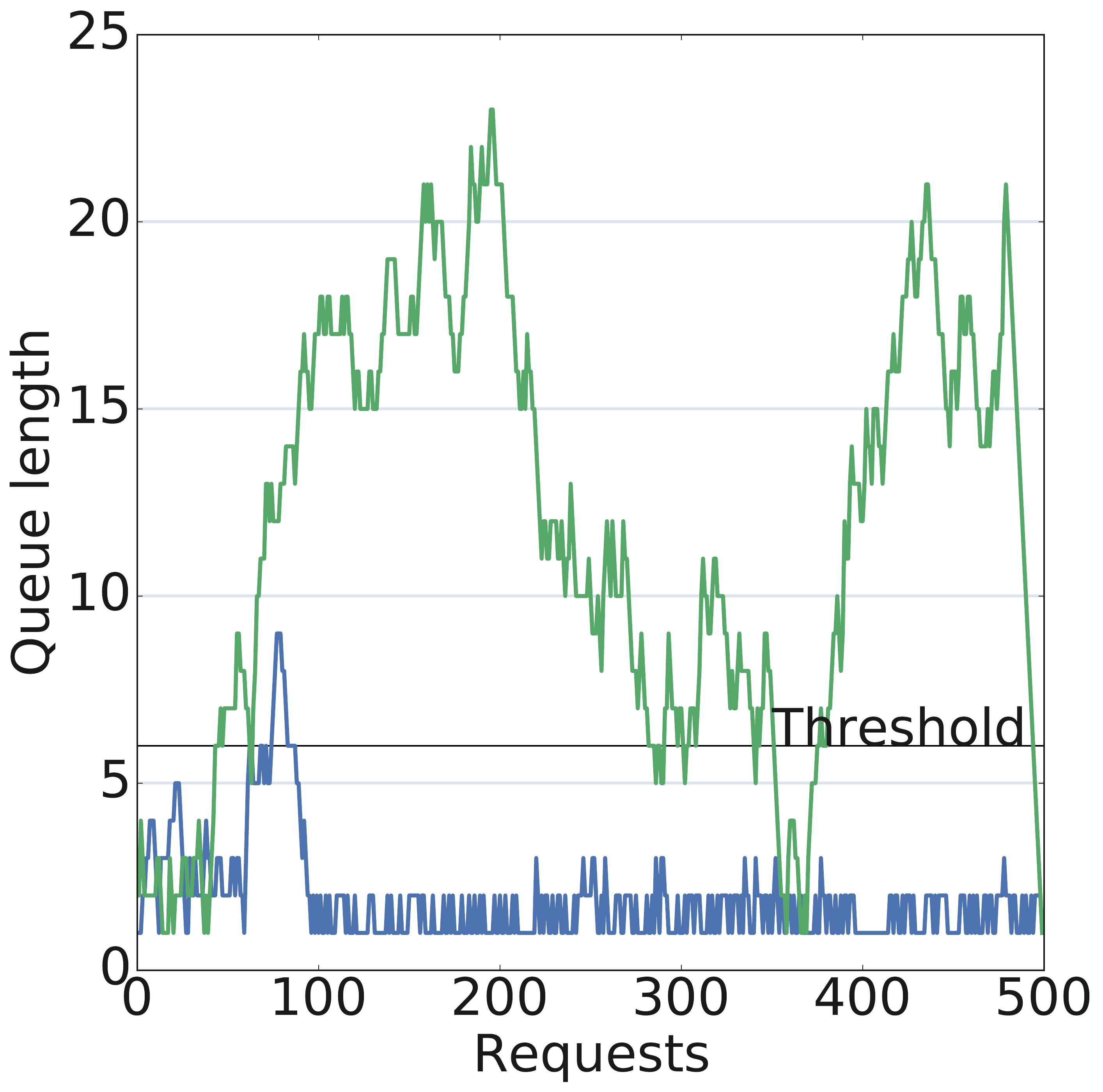}
	}
	\subfigure[Queue length of image upscaling with $\lambda = 17$ and a variant switch from psnr-large to psnr-small]{
		\label{fig:evaluation_variant_switch_image_upscaling_17_V2_V3_QL_small}
		\includegraphics[height=0.227\textwidth]{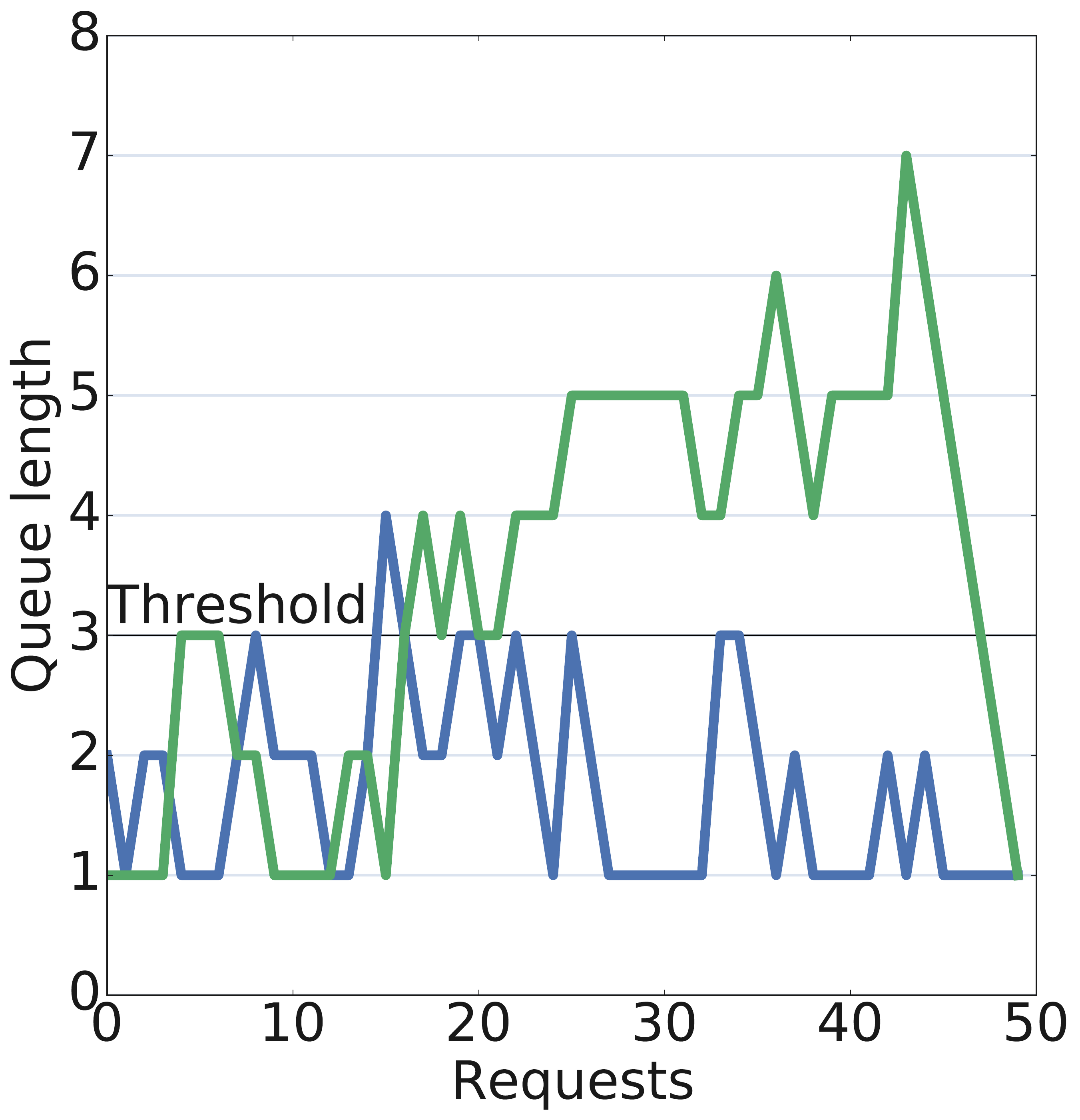}
	}
	\subfigure[Queue length of image upscaling with $\lambda = 17$ and a variant switch from gans to psnr-small]{
		\label{fig:evaluation_variant_switch_image_upscaling_17_V1_V3_QL_small}
		\includegraphics[height=0.227\textwidth]{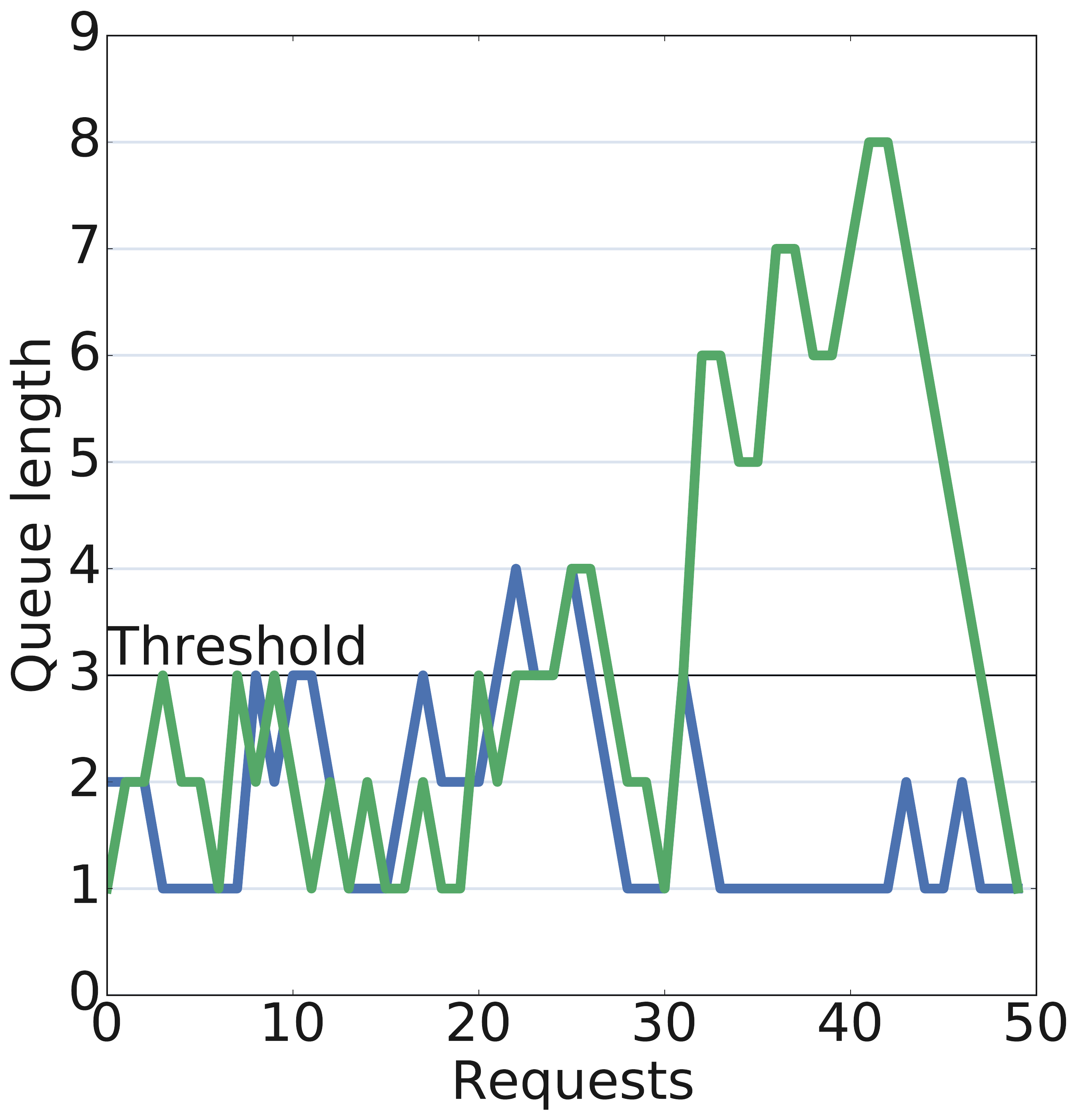}
	}
	\subfigure[Execution time of face detection with $\lambda = 15$]{
		\label{fig:evaluation_variant_switch_face_detection_15_ET_small}
		\includegraphics[width=0.225\textwidth]{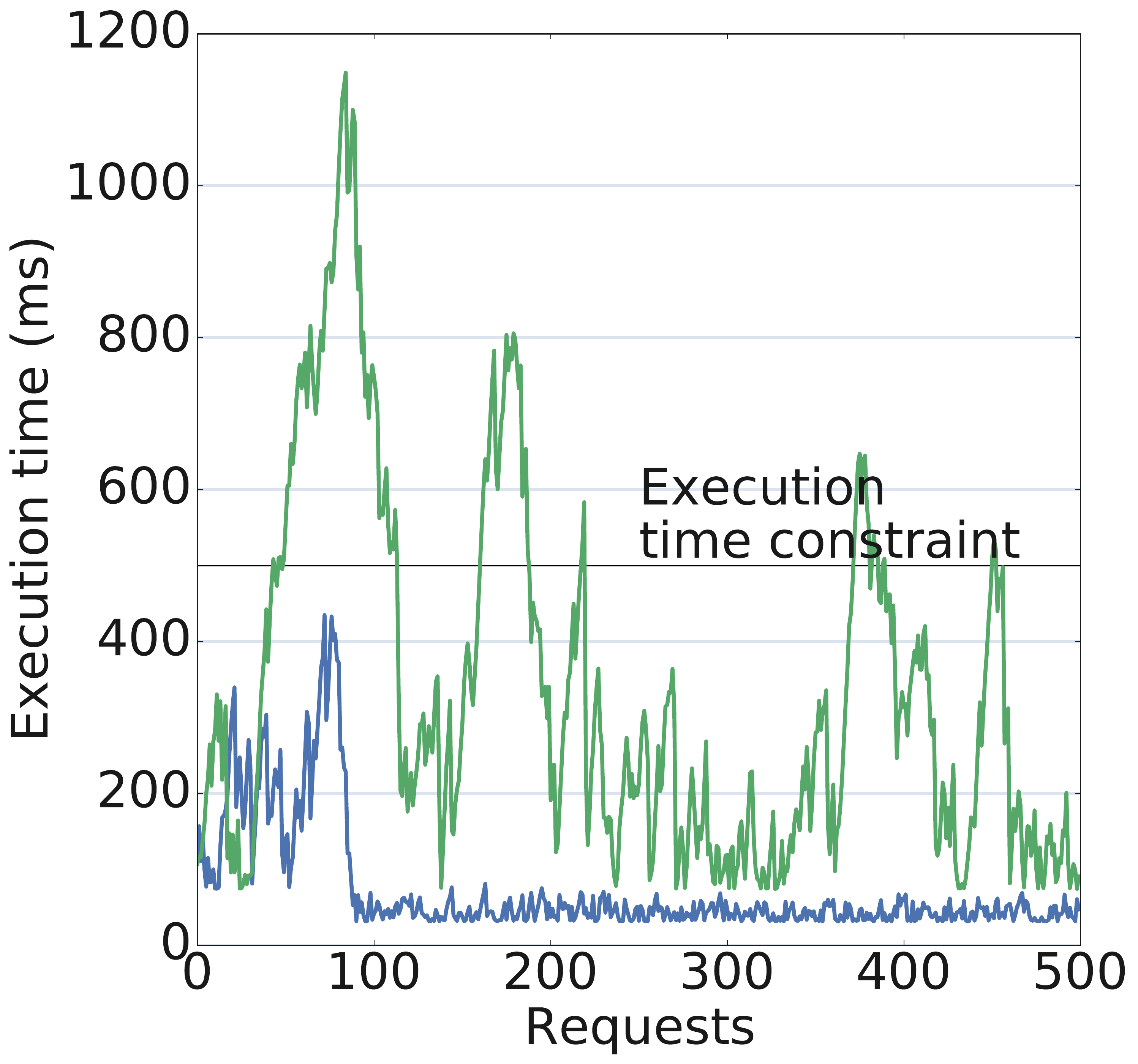}
	}
	\subfigure[Execution time of face detection with $\lambda = 17$]{
		\label{fig:evaluation_variant_switch_face_detection_17_ET_small}
		\includegraphics[width=0.225\textwidth]{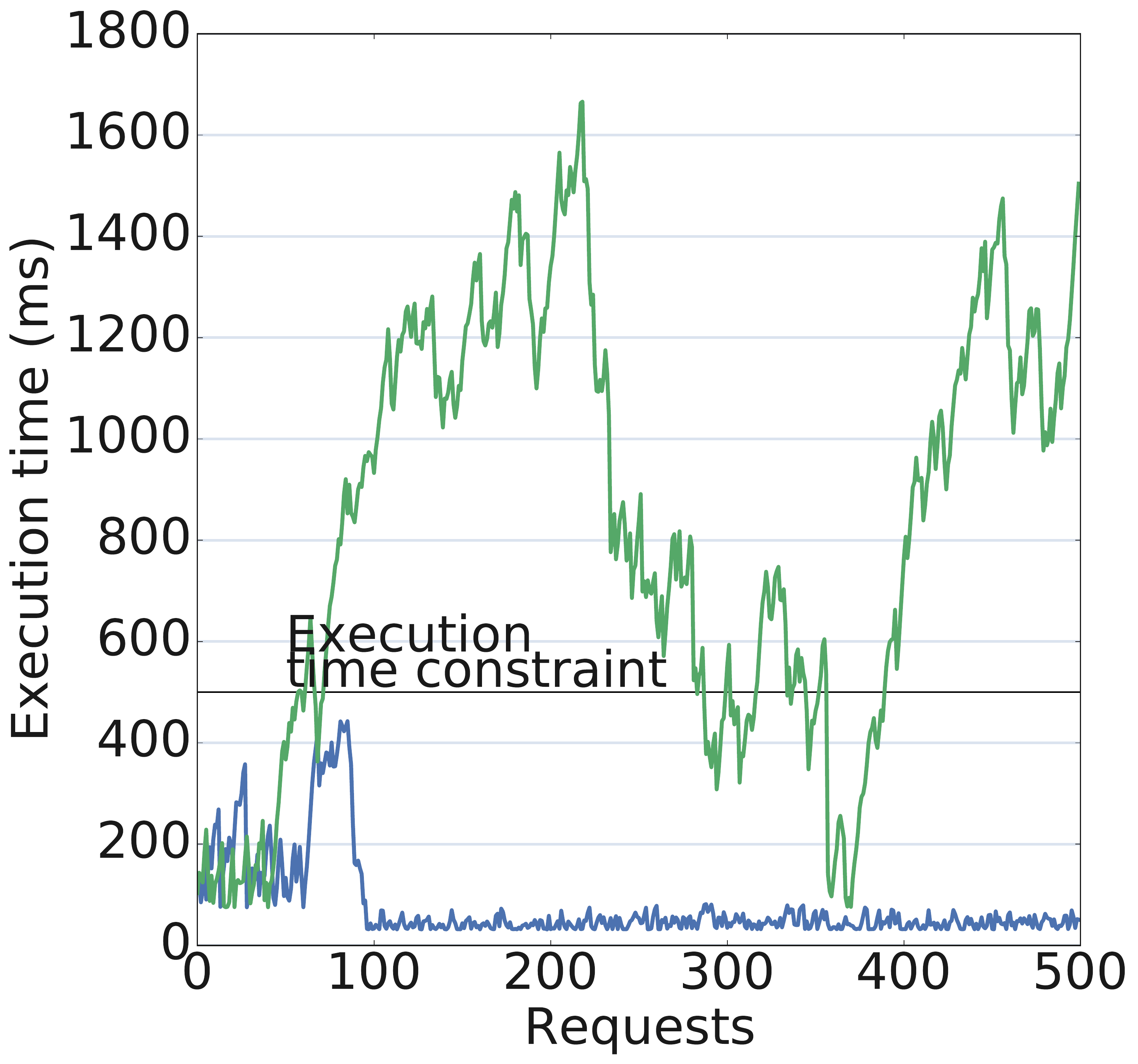}
	}
	\subfigure[Execution time of image upscaling with $\lambda = 17$ and a variant switch from psnr-large to psnr-small]{
		\label{fig:evaluation_variant_switch_image_upscaling_17_V2_V3_ET_small}
		\includegraphics[width=0.225\textwidth]{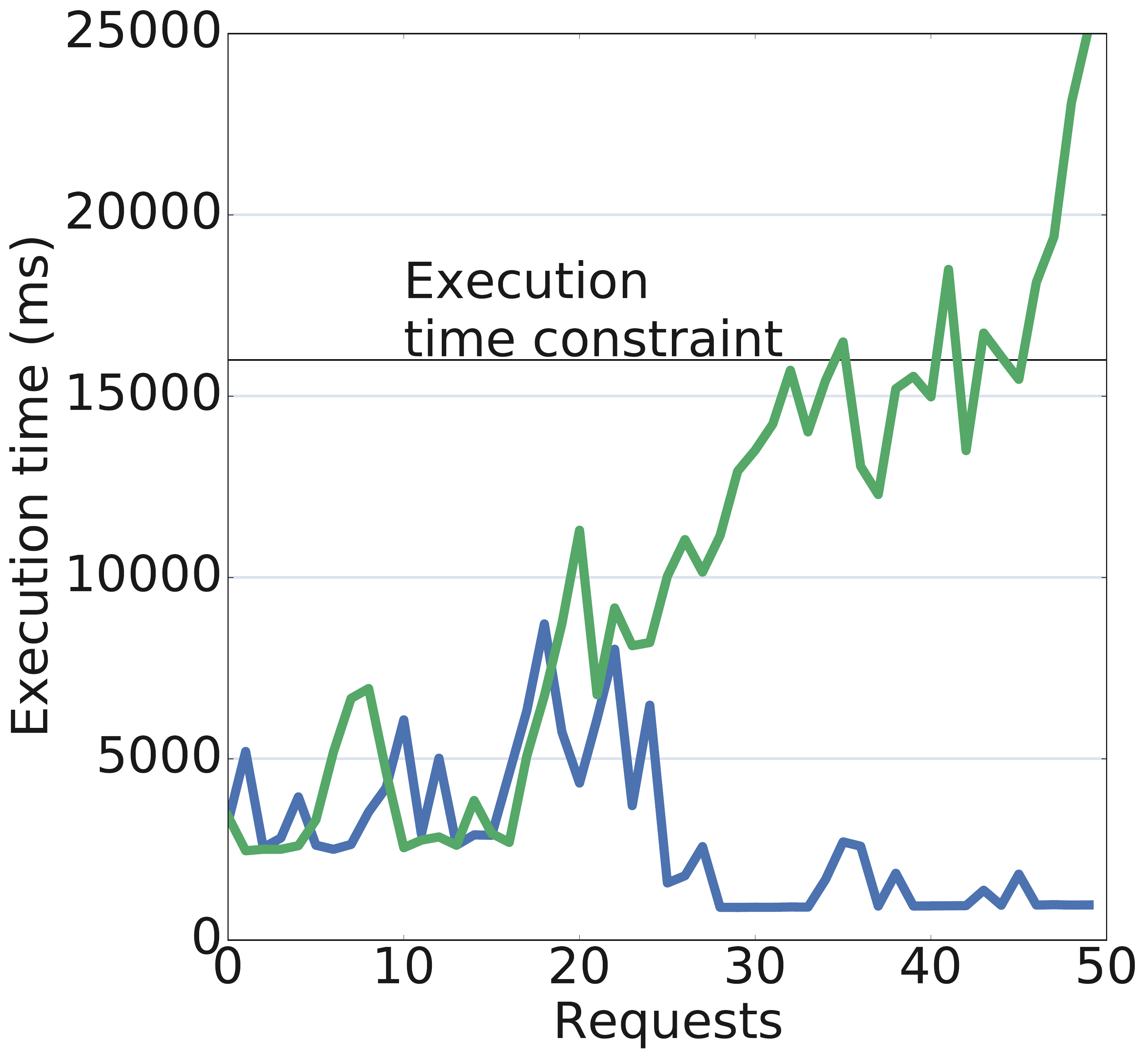}
	}
	\subfigure[Execution time of image upscaling with $\lambda = 17$ and a variant switch from gans to psnr-small]{
		\label{fig:evaluation_variant_switch_image_upscaling_17_V1_V3_ET_small}
		\includegraphics[width=0.225\textwidth]{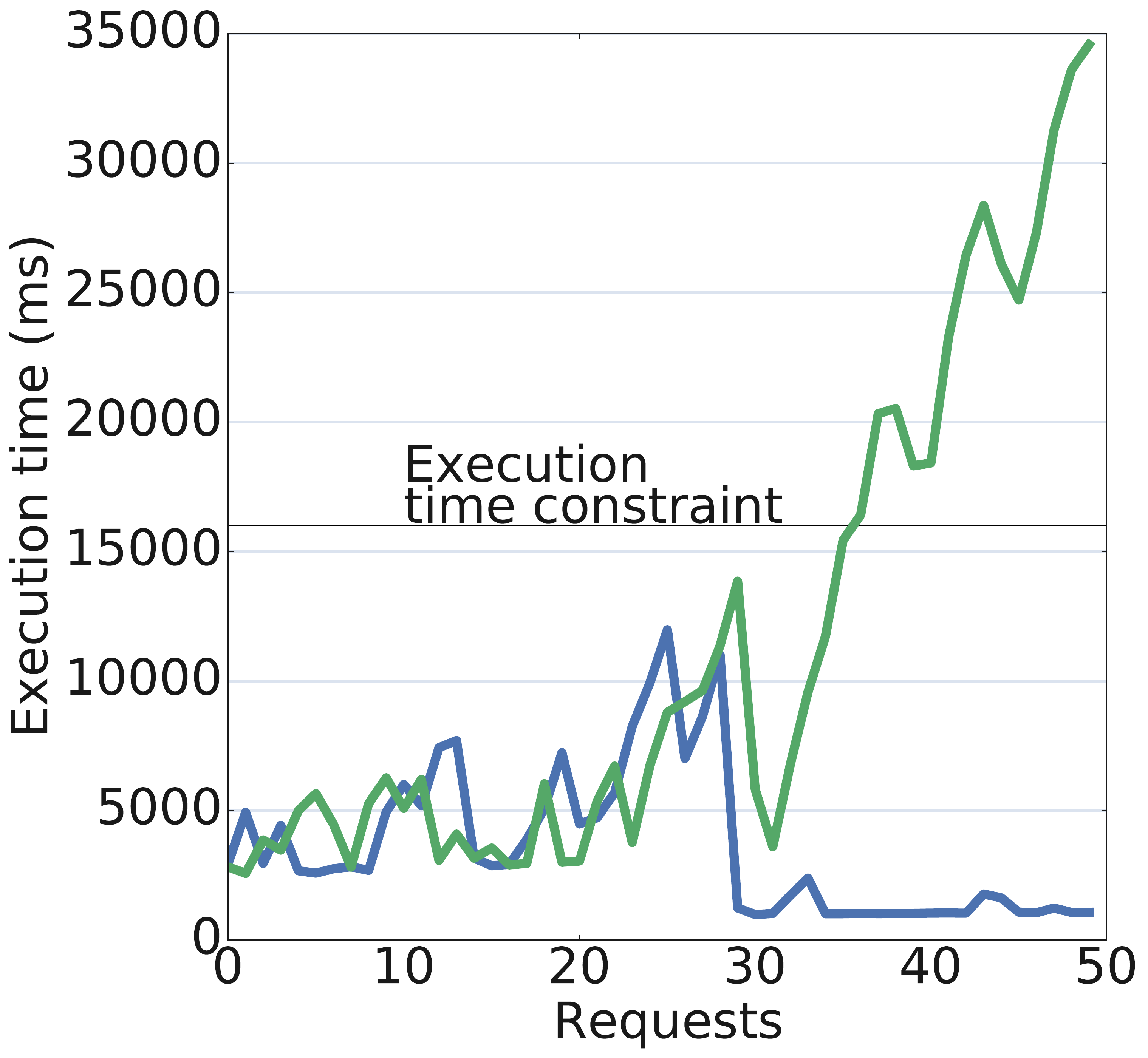}
	}

	\caption[Queue length and execution time of face detection and image upscaling service variants]{Queue length and execution time of face detection and image upscaling service variants}
	\label{fig:evaluation_variant_switch_small}
\end{figure*}

\paragraph*{Methodology and experimental setup} For our experiments, we send a number of requests via a request generator to the request queue of a running microservice.
The times between requests are drawn from an exponential distribution with parameter $\lambda$.
We use two of our microservices described in \Cref{subsec:implementation_microservices} for our experiments, namely face detection and image upscaling.
We evaluate our variant switching strategy with different $\lambda$ and measure the execution time of each request and the length of the request queue over the course of the experiments.
We compare our results with the baseline queue lengths and execution times of microservice executions where the variant switching is disabled.
The microservices are executed on a Lenovo ThinkCentre M920X Tiny (Intel Core i7-8700, 16\,GB RAM, Ubuntu 18.04). Such small scale, energy-efficient, yet powerful devices are representative examples of possible edge resources, e.g., when publicly deployed in urban environments~\cite{gedeon2018-sec}

\paragraph*{Experimental results}

\Cref{fig:evaluation_variant_switch_face_detection_15_QL_small,fig:evaluation_variant_switch_face_detection_17_QL_small,fig:evaluation_variant_switch_face_detection_15_ET_small,fig:evaluation_variant_switch_face_detection_17_ET_small} show the results of our variant switching experiments with the face detection microservice.
In our experiments, the face detection microservice changes its algorithm if a variant switch is performed.
The microservice starts with the \emph{Haar-Classifier} and eventually switches to the \emph{LBP-Classifier} as an algorithm for the face detection.
We conduct two variants of the experiment for the face detection microservice.
In each, we send 500 requests, define an execution time constraint of \unit[500]{ms}, and set the dampening parameter to $\alpha=1$.
This results in a variant switching threshold of 6 requests in the queue.
For the first experiment, we set the arrival rate to $\lambda = 15$, while for the second experiment, we increase the rate to $\lambda = 17$, so that the inter-arrival times of requests are shorter than in the first experiment.
As a baseline for comparison, we also report on the results when there is no switching of microservice variants.
In the experiment with  $\lambda=17$, the stability check in our switching strategy is disabled.
Otherwise, the variant that uses the \emph{Haar-Classifier} would not be chosen at all and a variant switch would not be performed.
\par
\Cref{fig:evaluation_variant_switch_face_detection_15_QL_small,fig:evaluation_variant_switch_face_detection_15_ET_small} show that as the execution time reaches the constraint and the queue length reaches the threshold, a variant switch to the \emph{LBP-Classifier} is initiated at request number 77.
The continued execution with the faster variant avoids a violation of the user-given execution time constraint of \unit[500]{ms}.
Without variant switching, the constraint is violated repeatedly.
Similar results can be observed in \Cref{fig:evaluation_variant_switch_face_detection_17_QL_small,fig:evaluation_variant_switch_face_detection_17_ET_small}.
Because of a greater $\lambda$ compared to the first experiment, the execution time and queue length grow faster.
This leads to a variant switch at request number 82.
Without variant switching, the microservice will be overwhelmed with requests, which leads to a steady increase in both the execution time and queue length.
As a consequence, the execution time constraint is violated for most of the requests after the first violation.

\Cref{fig:evaluation_variant_switch_image_upscaling_17_V2_V3_QL_small,fig:evaluation_variant_switch_image_upscaling_17_V1_V3_QL_small,fig:evaluation_variant_switch_image_upscaling_17_V2_V3_ET_small,fig:evaluation_variant_switch_image_upscaling_17_V1_V3_ET_small}
show the results of our variant switching experiments with the image upscaling microservice.
In these experiments, the microservice switches its variant by changing the pre-trained model that it uses.
For our experiments, we used the models \emph{psnr-small}, \emph{psnr-large}, and \emph{gans}.
Of these three models, \emph{psnr-small} leads to the shortest execution time while \emph{gans} leads to the longest.
We conducted our experiments with the image upscaling microservice two times.
In each, we send 50 requests, define an execution time constraint of \unit[16000]{ms}, and set the dampening parameter to $\alpha = 1$.
For the models \emph{gans} and \emph{psnr-large}, this leads to a threshold for the variant switching of 3.
We use values of $\lambda = 17$ for both experiments.
Again, we also include a baseline where no switch is performed for comparison.

\Cref{fig:evaluation_variant_switch_image_upscaling_17_V2_V3_QL_small,fig:evaluation_variant_switch_image_upscaling_17_V2_V3_ET_small} show the experimental results of the image upscaling microservice.
Because of the request arrival rate of $\lambda = 17$, the execution starts with the model \emph{psnr-large} instead of \emph{gans} and later switches to the model \emph{psnr-small}.
This decision is made because of the queue stability condition, so that an unstable request queue, where requests accumulate, is avoided.
Without the variant switch, the constraint would be violated for several requests.
\Cref{fig:evaluation_variant_switch_image_upscaling_17_V1_V3_QL_small,fig:evaluation_variant_switch_image_upscaling_17_V1_V3_ET_small} show the results of our second experiment with the image upscaling microservice and $\lambda = 17$.
For this experiment, we disable the aforementioned queue stability condition, so that we can enforce the execution of the microservice variant with the \emph{gans} model, leading to an unstable request queue.
At request number 29 a variant switch is performed to the variant with the \emph{psnr-small} model.
This model has been chosen according to our condition, that the average waiting time (\Cref{eq:avg_wait_time}) for requests has to be lower than the execution time constraint.
After the switch, constraint violations are avoided.
Without the switch, the request queue would fill up which leads to constant violations of the execution time constraint.

Additionally, we compare our variant switching approach to an alternative approach, where a service would be restarted in the variant it is supposed to switch to.
The process of restarting takes about \unit[1.8]{s} for the face detection service and \unit[4]{s} for the image upscaling service.
The variant switch approach we use in our experiments takes about \unit[17]{ms} for both the face detection and image upscaling microservices.

\paragraph*{Summary} We have shown that switching microservice variants at runtime can help ensuring execution time constraints.
We demonstrated this with two microservices and a threshold-based switching strategy.
It automatically decides when to switch the microservice variant and also decides which variant to switch to.
This approach avoids expensive re-starts of services in \EC environments.
Note that we plan to extend our switching strategy, so that switches to slower variants that deliver better results are performed when the queue has been stable enough for a sufficient amount of time.
This way, execution time constraints can be met while also maximizing the QoR.

\section{Discussion \& Future Work}
\label{sec:discussion_futurework}

To demonstrate the benefits of adaptable microservices and present our switching strategy (see the previous \Cref{sec:evaluation}), we implemented a subset of the overall design shown in \Cref{fig:implementation_framework}.
However, a number of practical issues remain for the global orchestration of service variants.
Our work opens up the following questions for future work:\\
\bpara{Hierarchical monitoring and control.} To ensure that application-specific constraints \WRT the execution time and result quality are met, the execution of service chains needs to be monitored.
Given the highly distributed nature of edge cloudlets, having only one centralized controller does not meet the scalability requirements of \EC.
Hence, we envision hierarchical monitoring and control mechanisms. \\
\bpara{Network control layer.} In a distributed \EC system, not only the resources on the edge nodes and the microservices' complexity influence the execution time but also the network conditions and types of connections between the nodes (e.g., when the microservices of one service chain run on different nodes).
Future work should take this into consideration in two aspects: First, fine-grained monitoring of network conditions can help in making runtime decisions for the placement and assignment of microservices.
Second, we can extend the control itself to the network layer, e.g., by reserving bandwidth on links or using SDN to control the data flow between edge nodes.\\
\bpara{Defining and weighting multiple QoR metrics.} As we have noted, the quality of a computation can be defined in different ways.
However, the interplay between user-perceived QoR and mathematical metrics for QoR is not well understood yet.
It also remains unclear how both types of QoR should be weighted if they are part of one service chain.
Furthermore, if a microservice instance is part of multiple service chains, this could lead to conflicts w.r.t. the individual optimization targets.
\\
\bpara{Defining service variants through SPLs.} Software product lines allow for a more general modeling of application variants.
Using this technique would also make it possible to model more complex dependencies between variants (e.g., when certain combinations of variants are mutually exclusive).

\section{Conclusion}
\label{sec:conclusion}

Based on three properties of \EC and its applications---constrained resources, tight constraints on the execution time, and flexibility regarding the quality of the computations---this paper proposed the general concept of \emph{adaptable microservices}.
Specifically, we defined microservices to be adaptable in three aspects, related to the \emph{internal functioning} of the microservices.
We designed the concept for the integration of adaptable microservices into an \EC framework.
\par
After having shown that we can accurately profile their execution times, we demonstrated the practical impact of adaptable microservice variants in relevant application domains of computer vision and image processing.
Adaptable microservices allow trading the quality of computations for lower resource utilization (manifested for example in a reduced execution time).
We furthermore demonstrated how switching service variants at runtime can help adapting to changing request patterns.
The proposed concept of microservice variants can help in mitigating the limited elasticity of \EC by \emph{adapting the services to the limitations of the execution infrastructure} and not vice versa.

\begin{small}
\section*{Acknowledgement}
This work has been cofunded by the German Research Foundation (DFG) and the National Nature Science Foundation of China (NSFC) joint project under Grant No. 392046569 (DFG) and No. 61761136014 (NSFC), and as part of the Collaborative Research Center 1053 - MAKI (DFG).
This work was supported by the \emph{AWS Cloud Credits for Research} program.
\end{small}

\bibliographystyle{IEEEtran}
\bibliography{refs}

\begin{thebibliography}{10}
\providecommand{\url}[1]{#1}
\csname url@samestyle\endcsname
\providecommand{\newblock}{\relax}
\providecommand{\bibinfo}[2]{#2}
\providecommand{\BIBentrySTDinterwordspacing}{\spaceskip=0pt\relax}
\providecommand{\BIBentryALTinterwordstretchfactor}{4}
\providecommand{\BIBentryALTinterwordspacing}{\spaceskip=\fontdimen2\font plus
\BIBentryALTinterwordstretchfactor\fontdimen3\font minus
  \fontdimen4\font\relax}
\providecommand{\BIBforeignlanguage}[2]{{%
\expandafter\ifx\csname l@#1\endcsname\relax
\typeout{** WARNING: IEEEtran.bst: No hyphenation pattern has been}%
\typeout{** loaded for the language `#1'. Using the pattern for}%
\typeout{** the default language instead.}%
\else
\language=\csname l@#1\endcsname
\fi
#2}}
\providecommand{\BIBdecl}{\relax}
\BIBdecl

\bibitem{gedeon2019-survey}
J.~Gedeon, F.~Brandherm, R.~Egert, T.~Grube, and M.~M{\"u}hlh{\"a}user, ``{What
  the Fog? Edge Computing Revisited: Promises, Applications and Future
  Challenges},'' \emph{IEEE Access}, vol.~7, pp. 152\,847--152\,878, 2019.

\bibitem{Shi2016a}
W.~Shi, J.~Cao, Q.~Zhang, Y.~Li, and L.~Xu, ``{Edge Computing: Vision and
  Challenges},'' \emph{{IEEE} Internet of Things Journal}, vol.~3, no.~5, pp.
  637--646, 2016.

\bibitem{Satyanarayanan17-Emergence}
M.~Satyanarayanan, ``{The Emergence of Edge Computing},'' \emph{{IEEE}
  Computer}, vol.~50, no.~1, pp. 30--39, 2017.

\bibitem{Pahl2015-Containers}
C.~Pahl and B.~Lee, ``{Containers and Clusters for Edge Cloud Architectures - A
  Technology Review},'' in \emph{Proc. of the 3rd International Conference on
  Future Internet of Things and Cloud}, 2015, pp. 379--386.

\bibitem{Morabito2018a}
R.~Morabito, V.~Cozzolino, A.~Y. Ding, N.~Beijar, and J.~Ott, ``{Consolidate
  IoT Edge Computing with Lightweight Virtualization},'' \emph{{IEEE} Network},
  vol.~32, no.~1, pp. 102--111, 2018.

\bibitem{Kaur2017-Energy}
K.~Kaur, T.~Dhand, N.~Kumar, and S.~Zeadally, ``{Container-as-a-Service at the
  Edge: Trade-off between Energy Efficiency and Service Availability at Fog
  Nano Data Centers},'' \emph{{IEEE} Wireless Communnications}, vol.~24, no.~3,
  pp. 48--56, 2017.

\bibitem{Liu2016-Paradrop}
P.~Liu, D.~Willis, and S.~Banerjee, ``{ParaDrop: Enabling Lightweight
  Multi-tenancy at the Network's Extreme Edge},'' in \emph{Proc. of the
  {IEEE/ACM} Symposium on Edge Computing (SEC)}, 2016, pp. 1--13.

\bibitem{Filip2018-Microservices}
I.~Filip, F.~Pop, C.~Serbanescu, and C.~Choi, ``{Microservices Scheduling Model
  Over Heterogeneous Cloud-Edge Environments As Support for IoT
  Applications},'' \emph{{IEEE} Internet of Things Journal}, vol.~5, no.~4, pp.
  2672--2681, 2018.

\bibitem{Alam2018-Microservices}
M.~Alam, J.~Rufino, J.~Ferreira, S.~H. Ahmed, N.~Shah, and Y.~Chen,
  ``{Orchestration of Microservices for IoT Using Docker and Edge Computing},''
  \emph{{IEEE} Communications Magazine}, vol.~56, no.~9, pp. 118--123, 2018.

\bibitem{Salaht2020-ServicePlacement}
F.~A. Salaht, F.~Desprez, and A.~Lebre, ``{An Overview of Service Placement
  Problem in Fog and Edge Computing},'' \emph{ACM Computing Surveys}, vol.~53,
  no.~3, Jun. 2020.

\bibitem{Ouyang2018-FollowMeEdge}
T.~Ouyang, Z.~Zhou, and X.~Chen, ``{Follow Me at the Edge: Mobility-Aware
  Dynamic Service Placement for Mobile Edge Computing},'' \emph{{IEEE} JSAC.},
  vol.~36, no.~10, pp. 2333--2345, 2018.

\bibitem{Wang2018-ServiceMigration}
S.~Wang, J.~Xu, N.~Zhang, and Y.~Liu, ``A survey on service migration in mobile
  edge computing,'' \emph{{IEEE} Access}, vol.~6, pp. 23\,511--23\,528, 2018.

\bibitem{Ma2017-ServiceHandoff}
L.~Ma, S.~Yi, and Q.~Li, ``{Efficient service handoff across edge servers via
  docker container migration},'' in \emph{Proc. of the {IEEE/ACM} Symposium on
  Edge Computing (SEC)}, J.~Zhang, M.~Chiang, and B.~M. Maggs, Eds., 2017, pp.
  11:1--11:13.

\bibitem{Chippa2013-ACResilience}
V.~K. Chippa, S.~T. Chakradhar, K.~Roy, and A.~Raghunathan, ``{Analysis and
  characterization of inherent application resilience for approximate
  computing},'' in \emph{Proc. of the 50th Annual Design Automation Conference
  (DAC)}, 2013, pp. 113:1--113:9.

\bibitem{Satyanarayanan2009-Cloudlets}
M.~Satyanarayanan, P.~Bahl, R.~C\'aceres, and N.~Davies, ``{The Case for
  VM-Based Cloudlets in Mobile Computing},'' \emph{{IEEE} Pervasive Computing},
  vol.~8, no.~4, pp. 14--23, 2009.

\bibitem{Dolui2017-EdgeImplementations}
K.~Dolui and S.~K. Datta, ``{Comparison of edge computing implementations: Fog
  computing, cloudlet and mobile edge computing},'' in \emph{Proc. of the
  Global Internet of Things Summit (GIoTS)}, 2017, pp. 1--6.

\bibitem{Carrega2017-MEC}
A.~Carrega, M.~Repetto, P.~Gouvas, and A.~Zafeiropoulos, ``{A Middleware for
  Mobile Edge Computing},'' \emph{{IEEE} Cloud Computing}, vol.~4, no.~4, pp.
  26--37, 2017.

\bibitem{Ferrer2019-AdHocEdge}
A.~J. Ferrer, J.~M. Marqu{\`{e}}s, and J.~Jorba, ``{Ad-Hoc Edge Cloud: {A}
  Framework for Dynamic Creation of Edge Computing Infrastructures},'' in
  \emph{Proc. of the 28th International Conference on Computer Communication
  and Networks (ICCCN)}, 2019, pp. 1--7.

\bibitem{Wang2017-OnlinePlacement}
S.~Wang, M.~Zafer, and K.~K. Leung, ``{Online Placement of Multi-Component
  Applications in Edge Computing Environments},'' \emph{{IEEE} Access}, vol.~5,
  pp. 2514--2533, 2017.

\bibitem{Chen2019-ServiceMigration}
M.~Chen, W.~Li, G.~Fortino, Y.~Hao, L.~Hu, and I.~Humar, ``A dynamic service
  migration mechanism in edge cognitive computing,'' \emph{{ACM} Trans.
  Internet Techn.}, vol.~19, no.~2, pp. 30:1--30:15, 2019.

\bibitem{Pamboris2015-NOMAD}
A.~Pamboris, M.~Baguena, A.~L. Wolf, P.~Manzoni, and P.~R. Pietzuch, ``{Demo:
  NOMAD: An Edge Cloud Platform for Hyper-Responsive Mobile Apps},'' in
  \emph{Proc. of the 13th Annual International Conference on Mobile Systems,
  Applications, and Services (MobiSys)}, 2015, p. 459.

\bibitem{Mortazavi2017-Cloudpath}
S.~H. Mortazavi, M.~Salehe, C.~S. Gomes, C.~Phillips, and E.~de~Lara,
  ``{Cloudpath: a multi-tier cloud computing framework},'' in \emph{Proc. of
  the {ACM/IEEE} Symposium on Edge Computing (SEC)}, 2017, pp. 20:1--20:13.

\bibitem{Bhardwaj2016-AirBox}
K.~{Bhardwaj}, M.~{Shih}, P.~{Agarwal}, A.~{Gavrilovska}, T.~{Kim}, and
  K.~{Schwan}, ``{Fast, Scalable and Secure Onloading of Edge Functions Using
  AirBox},'' in \emph{Proc. of the IEEE/ACM Symposium on Edge Computing (SEC)},
  2016, pp. 14--27.

\bibitem{Wang2019-ScalableEdge}
J.~Wang, Z.~Feng, S.~A. George, R.~Iyengar, P.~Pillai, and M.~Satyanarayanan,
  ``{Towards scalable edge-native applications},'' in \emph{Proc. of the
  {ACM/IEEE} Symposium on Edge Computing (SEC)}, 2019, pp. 152--165.

\bibitem{Fowler2012-Microservices}
M.~Fowler, ``{Microservices - a definition of this new architectural term},''
  \url{https://martinfowler.com/articles/microservices.html}, 2014, accessed:
  2021-01-07.

\bibitem{Balalaie2016-Microservices}
A.~Balalaie, A.~Heydarnoori, and P.~Jamshidi, ``{Microservices Architecture
  Enables DevOps: Migration to a Cloud-Native Architecture},'' \emph{{IEEE}
  Software}, vol.~33, no.~3, pp. 42--52, 2016.

\bibitem{Kang2016-DevOps}
H.~{Kang}, M.~{Le}, and S.~{Tao}, ``{Container and Microservice Driven Design
  for Cloud Infrastructure DevOps},'' in \emph{Proc. of the 2016 IEEE
  International Conference on Cloud Engineering (IC2E)}, 2016, pp. 202--211.

\bibitem{Jamshidi2018-Microservices}
P.~{Jamshidi}, C.~{Pahl}, N.~C. {Mendonça}, J.~{Lewis}, and S.~{Tilkov},
  ``{Microservices: The Journey So Far and Challenges Ahead},'' \emph{IEEE
  Software}, vol.~35, no.~3, pp. 24--35, 2018.

\bibitem{Dragoni2018-MicroserviceScalability}
N.~Dragoni, I.~Lanese, S.~T. Larsen, M.~Mazzara, R.~Mustafin, and L.~Safina,
  ``{Microservices: How To Make Your Application Scale},'' in \emph{Proc. of
  the 11th International Andrei P. Ershov Informatics Conference (PSI)}, 2018,
  pp. 95--104.

\bibitem{Taibi2017-Microservices}
D.~Taibi, V.~Lenarduzzi, and C.~Pahl, ``{Processes, Motivations, and Issues for
  Migrating to Microservices Architectures: An Empirical Investigation},''
  \emph{{IEEE} Cloud Computing}, vol.~4, no.~5, pp. 22--32, 2017.

\bibitem{Soldani2018-Microservices}
J.~Soldani, D.~A. Tamburri, and W.~van~den Heuvel, ``{The pains and gains of
  microservices: A Systematic grey literature review},'' \emph{Journal of
  Systems and Software}, vol. 146, pp. 215--232, 2018.

\bibitem{Dragoni2017-Microservices}
N.~Dragoni, S.~Giallorenzo, A.~Lluch{-}Lafuente, M.~Mazzara, F.~Montesi,
  R.~Mustafin, and L.~Safina, ``{Microservices: Yesterday, Today, and
  Tomorrow},'' in \emph{Present and Ulterior Software Engineering}, 2017, pp.
  195--216.

\bibitem{Hassan2017-MicroserviceGranularity}
S.~{Hassan}, N.~{Ali}, and R.~{Bahsoon}, ``{Microservice Ambients: An
  Architectural Meta-Modelling Approach for Microservice Granularity},'' in
  \emph{Proc.of the 2017 IEEE International Conference on Software Architecture
  (ICSA)}, 2017, pp. 1--10.

\bibitem{Hassan2016-MicroserviceDesign}
S.~Hassan and R.~Bahsoon, ``{Microservices and Their Design Trade-Offs: A
  Self-Adaptive Roadmap},'' in \emph{Proc. of the {IEEE} International
  Conference on Services Computing (SCC)}, 2016, pp. 813--818.

\bibitem{Papazoglou2003-SOA}
M.~P. Papazoglou, ``{Service-Oriented Computing: Concepts, Characteristics and
  Directions},'' in \emph{Proc. of the Fourth International Conference on Web
  Information Systems Engineering}, ser. WISE ’03, 2003, pp. 3--12.

\bibitem{Chang2007-ServiceAdaptation}
S.~H. {Chang}, H.~J. {La}, and S.~D. {Kim}, ``{A Comprehensive Approach to
  Service Adaptation},'' in \emph{Proc. of the IEEE International Conference on
  Service-Oriented Computing and Applications (SOCA '07)}, 2007, pp. 191--198.

\bibitem{Hirschfeld2006-DynamicAdaptation}
R.~Hirschfeld and K.~Kawamura, ``{Dynamic service adaptation},'' \emph{Software
  Practice and Experience}, vol.~36, no. 11-12, pp. 1115--1131, 2006.

\bibitem{McGregor2002-SPL}
J.~D. McGregor, L.~M. Northrop, S.~Jarrad, and K.~Pohl, ``{Guest Editors'
  Introduction: Initiating Software Product Lines},'' \emph{{IEEE} Software},
  vol.~19, no.~4, pp. 24--27, 2002.

\bibitem{VanGurp2001-SPLVariation}
J.~{van Gurp}, J.~{Bosch}, and M.~{Svahnberg}, ``{On the notion of variability
  in software product lines},'' in \emph{Proc. of the Working IEEE/IFIP
  Conference on Software Architecture}, 2001, pp. 45--54.

\bibitem{Beuche2007-SPLFM}
D.~Beuche and M.~Dalgarno, ``{Software product line engineering with feature
  models},'' \emph{Overload Journal}, vol.~78, pp. 5--8, 2007.

\bibitem{Sanchez2013-Featuremodels}
L.~E. Sanchez, S.~Moisan, and J.~Rigault, ``{Metrics on feature models to
  optimize configuration adaptation at run time},'' in \emph{Proc. of the 1st
  International Workshop on Combining Modelling and Search-Based Software
  Engineering, (CMSBSE)}, 2013, pp. 39--44.

\bibitem{Hallsteinsen2008-DSPL}
S.~{Hallsteinsen}, M.~{Hinchey}, S.~{Park}, and K.~{Schmid}, ``{Dynamic
  Software Product Lines},'' \emph{Computer}, vol.~41, no.~4, pp. 93--95, 2008.

\bibitem{Weckesser2018-SPL}
M.~Weckesser, R.~Kluge, M.~Pfannem{\"{u}}ller, M.~Matth{\'{e}},
  A.~Sch{\"{u}}rr, and C.~Becker, ``{Optimal reconfiguration of dynamic
  software product lines based on performance-influence models},'' in
  \emph{Proc. of the 22nd International Systems and Software Product Line
  Conference (SPLC)}, 2018, pp. 98--109.

\bibitem{Gholami2019-DockerMV}
S.~Gholami, A.~Goli, C.-P. Bezemer, and H.~Khazaei, ``{A Framework for
  Satisfying the Performance Requirements of Containerized Software Systems
  Through Multi-Versioning},'' in \emph{Proc. of the International Conference
  on Performance Engineering (ICPE)}, 2019, pp. 1--11.

\bibitem{Kannan2019-GrandSLAM}
R.~S. Kannan, L.~Subramanian, A.~Raju, J.~Ahn, J.~Mars, and L.~Tang,
  ``{GrandSLAm: Guaranteeing SLAs for Jobs in Microservices Execution
  Frameworks},'' in \emph{Proc. of the 14th EuroSys Conference}, 2019, pp.
  34:1--34:16.

\bibitem{Mendonca2018-Adaptation}
N.~C. Mendon{\c{c}}a, D.~Garlan, B.~R. Schmerl, and J.~C{\'{a}}mara,
  ``{Generality vs. reusability in architecture-based self-adaptation: The case
  for self-adaptive microservices},'' in \emph{Proc. of the 12th European
  Conference on Software Architecture: Companion Proceedings, ({ECSA})}, 2018,
  pp. 18:1--18:6.

\bibitem{Bhattacharya2017-OffloadingAdaptation}
A.~Bhattacharya and P.~De, ``{A survey of adaptation techniques in computation
  offloading},'' \emph{Journal of Network and Computer Applications}, vol.~78,
  pp. 97--115, 2017.

\bibitem{Zhang2018-Stream}
B.~Zhang, X.~Jin, S.~Ratnasamy, J.~Wawrzynek, and E.~A. Lee, ``{AWStream:
  adaptive wide-area streaming analytics},'' in \emph{Proc. of the 2018
  Conference of the {ACM} Special Interest Group on Data Communication
  (SIGCOMM)}, 2018, pp. 236--252.

\bibitem{Zhou2015-IoTSA}
S.~{Zhou}, K.~{Lin}, J.~{Na}, C.~{Chuang}, and C.~{Shih}, ``{Supporting Service
  Adaptation in Fault Tolerant Internet of Things},'' in \emph{Proc. of the
  2015 IEEE 8th International Conference on Service-Oriented Computing and
  Applications (SOCA)}, 2015, pp. 65--72.

\bibitem{Mittal2016-ApproximateSurvey}
S.~Mittal, ``{A Survey of Techniques for Approximate Computing},'' \emph{{ACM}
  Computing Surveys}, vol.~48, no.~4, pp. 62:1--62:33, 2016.

\bibitem{Machidon2020-ApproxMC}
O.~Machidon, T.~Fajfar, and V.~Pejovi{\'c}, ``{Implementing Approximate Mobile
  Computing},'' in \emph{Proc. of the 2020 Workshop on Approximate Computing
  Across the Stack (WAX)}, 2020, pp. 1--3.

\bibitem{Moreau2018-Taxonomy}
T.~Moreau, J.~S. Miguel, M.~Wyse, J.~Bornholt, A.~Alaghi, L.~Ceze, N.~D.~E.
  Jerger, and A.~Sampson, ``{A Taxonomy of General Purpose Approximate
  Computing Techniques},'' \emph{Embedded Systems Letters}, vol.~10, no.~1, pp.
  2--5, 2018.

\bibitem{Gupta2011-ImpreciseAdders}
V.~Gupta, D.~Mohapatra, S.~P. Park, A.~Raghunathan, and K.~Roy, ``{IMPACT:
  imprecise adders for low-power approximate computing},'' in \emph{Proc. of
  the 2011 International Symposium on Low Power Electronics and Design
  (ISLPED)}, 2011, pp. 409--414.

\bibitem{Ye2013-ApproxAdders}
R.~Ye, T.~Wang, F.~Yuan, R.~Kumar, and Q.~Xu, ``{On reconfiguration-oriented
  approximate adder design and its application},'' in \emph{Proc. of the
  {IEEE/ACM} International Conference on Computer-Aided Design (ICCAD)}, 2013,
  pp. 48--54.

\bibitem{Mohapatra2011-VoltageScaling}
D.~Mohapatra, V.~K. Chippa, A.~Raghunathan, and K.~Roy, ``{Design of
  voltage-scalable meta-functions for approximate computing},'' in \emph{Proc.
  of the 2011 Design, Automation {\&} Test in Europe Conference {\&} Exhibition
  (DATE)}, 2011, pp. 950--955.

\bibitem{Guo2018-FoggyCache}
P.~Guo, B.~Hu, R.~Li, and W.~Hu, ``{FoggyCache: Cross-Device Approximate
  Computation Reuse},'' in \emph{Proc. of the 24th Annual International
  Conference on Mobile Computing and Networking (MobiCom)}, 2018, pp. 19--34.

\bibitem{Perez2017-Mapreduce}
J.~F. P{\'{e}}rez, R.~Birke, and L.~Y. Chen, ``{On the latency-accuracy
  tradeoff in approximate MapReduce jobs},'' in \emph{Proc. of the 2017 {IEEE}
  Conference on Computer Communications (INFOCOM)}, 2017, pp. 1--9.

\bibitem{Agrawal2016-AC}
A.~Agrawal, J.~Choi, K.~Gopalakrishnan, S.~Gupta, R.~Nair, J.~Oh, D.~A. Prener,
  S.~Shukla, V.~Srinivasan, and Z.~Sura, ``{Approximate computing: Challenges
  and opportunities},'' in \emph{Proc. of the {IEEE} International Conference
  on Rebooting Computing (ICRC)}, 2016, pp. 1--8.

\bibitem{Zhang2014-Approxit}
Q.~Zhang, F.~Yuan, R.~Ye, and Q.~Xu, ``{ApproxIt: An Approximate Computing
  Framework for Iterative Methods},'' in \emph{Proc. of the 51st Annual Design
  Automation Conference 2014 (DAC)}, 2014, pp. 97:1--97:6.

\bibitem{Almurib2018-Compression}
H.~A.~F. Almurib, T.~N. Kumar, and F.~Lombardi, ``{Approximate DCT Image
  Compression Using Inexact Computing},'' \emph{{IEEE} Transactions on
  Computers}, vol.~67, no.~2, pp. 149--159, 2018.

\bibitem{Zhang2015-ANN}
Q.~Zhang, T.~Wang, Y.~Tian, F.~Yuan, and Q.~Xu, ``{ApproxANN: an approximate
  computing framework for artificial neural network},'' in \emph{Proc. of the
  2015 Design, Automation {\&} Test in Europe Conference {\&} Exhibition
  (DATE)}, 2015, pp. 701--706.

\bibitem{Chen2018-DL}
C.~Chen, J.~Choi, K.~Gopalakrishnan, V.~Srinivasan, and S.~Venkataramani,
  ``{Exploiting approximate computing for deep learning acceleration},'' in
  \emph{Proc. of the 2018 Design, Automation {\&} Test in Europe Conference
  {\&} Exhibition (DATE)}, 2018, pp. 821--826.

\bibitem{Zamani2017-Edge}
A.~R. Zamani, I.~Petri, J.~D. Montes, O.~F. Rana, and M.~Parashar,
  ``{Edge-Supported Approximate Analysis for Long Running Computations},'' in
  \emph{Proc. of the 5th {IEEE} International Conference on Future Internet of
  Things and Cloud (FiCloud)}, 2017, pp. 321--328.

\bibitem{Wen2018-ApproxIoT}
Z.~Wen, D.~L. Quoc, P.~Bhatotia, R.~Chen, and M.~Lee, ``{ApproxIoT: Approximate
  Analytics for Edge Computing},'' in \emph{Proc. of the 38th {IEEE}
  International Conference on Distributed Computing Systems (ICDCS)}, 2018, pp.
  411--421.

\bibitem{Schaefer2016-QoC}
D.~Sch{\"{a}}fer, J.~Edinger, J.~M. Paluska, S.~VanSyckel, and C.~Becker,
  ``{Tasklets: "Better than Best-Effort" Computing},'' in \emph{Proc. of the
  25th International Conference on Computer Communication and Networks
  (ICCCN)}, 2016, pp. 1--11.

\bibitem{Edinger2017-Dev}
J.~Edinger, D.~Sch{\"{a}}fer, M.~Breitbach, and C.~Becker, ``{Developing
  distributed computing applications with Tasklets},'' in \emph{Proc. of the
  2017 {IEEE} International Conference on Pervasive Computing and
  Communications Workshops (PerCom Workshops)}, 2017, pp. 94--96.

\bibitem{Pejovic2018-MAC}
V.~Pejovic, ``{Towards Approximate Mobile Computing},'' \emph{GetMobile: Mobile
  Computing and Communications}, vol.~22, no.~4, pp. 9--12, 2018.

\bibitem{Sidiroglou2011-LoopPerforation}
S.~Sidiroglou{-}Douskos, S.~Misailovic, H.~Hoffmann, and M.~C. Rinard,
  ``{Managing performance vs. accuracy trade-offs with loop perforation},'' in
  \emph{Proc. of the SIGSOFT/FSE'11 19th {ACM} {SIGSOFT} Symposium on the
  Foundations of Software Engineering {(FSE-19)} and ESEC'11: 13th European
  Software Engineering Conference (ESEC-13)}, 2011, pp. 124--134.

\bibitem{Benavides2010-Featuremodels}
D.~Benavides, S.~Segura, and A.~R. Cort{\'{e}}s, ``{Automated analysis of
  feature models 20 years later: A literature review},'' \emph{Information
  Systems}, vol.~35, no.~6, pp. 615--636, 2010.

\bibitem{gedeon2019-microserviceOffloading}
J.~Gedeon, M.~Wagner, J.~Heuschkel, L.~Wang, and M.~M{\"u}hlh{\"a}user, ``{A
  Microservice Store for Efficient Edge Offloading},'' in \emph{Proc. of the
  {IEEE} Global Communications Conference (GLOBECOM)}, 2019, pp. 1--6.

\bibitem{gedeon2018-operatorPlacement}
J.~Gedeon, M.~Stein, L.~Wang, and M.~M{\"u}hlh{\"a}user, ``{On Scalable
  In-Network Operator Placement for Edge Computing},'' in \emph{Proc. of the
  27th International Conference on Computer Communication and Networks
  (ICCCN)}, 2018, pp. 1--9.

\bibitem{Gkioxari2019-3DMesh}
G.~Gkioxari, J.~Johnson, and J.~Malik, ``Mesh {R-CNN},'' in \emph{Proc. of the
  {IEEE/CVF} International Conference on Computer Vision (ICCV)}, 2019, pp.
  9784--9794.

\bibitem{Kadir2014-FaceDetectionClassifiers}
K.~Kadir, M.~K. Kamaruddin, H.~Nasir, S.~I. Safie, and Z.~A.~K. Bakti, ``{A
  comparative study between LBP and Haar-like features for Face Detection using
  OpenCV},'' in \emph{Proc. of the 4th International Conference on Engineering
  Technology and Technopreneuship (ICE2T)}, 2014, pp. 335--339.

\bibitem{gedeon2018-sec}
J.~Gedeon, M.~Stein, J.~Krisztinkovics, P.~Felka, K.~Keller, C.~Meurisch,
  L.~Wang, and M.~M{\"u}hlh{\"a}user, ``{From Cell Towers to Smart Street
  Lamps: Placing Cloudlets on Existing Urban Infrastructures},'' in \emph{Proc.
  of the {IEEE/ACM} Symposium on Edge Computing (SEC)}, 2018, pp. 187--202.

\end{thebibliography}
\end{document}